\documentclass[aps,pra,reprint,superscriptaddress]{revtex4-2}
\usepackage{amssymb,bm,mathrsfs,mathtools}
\usepackage{graphicx,tabularx}
\usepackage[dvipsnames]{xcolor}
\usepackage[bookmarks=false]{hyperref}
\hypersetup{colorlinks=true,allcolors=blue}

\newcommand{\ii}{\mkern1mu \mathrm{i}\mkern1mu}
\newcommand{\ee}{\mathrm{e}}
\newcommand{\dd}{\mathop{}\!\mathrm{d}}

\newcommand{\CZ}{\ensuremath{\mathrm{CZ}}}
\newcommand{\CZf}{\ensuremath{\mathsf{CZ}[\bm \phi]}}
\newcommand{\CNOT}{\ensuremath{\mathsf{CNOT}}}

\newcommand{\mr}[1]{\mathrm{#1}}
\newcommand{\mb}[1]{\mathbf{#1}}

\newcommand{\wt}[1]{\widetilde{#1}}
\newcommand{\uvec}[1]{\hat{\mathbf{#1}}}

\DeclareMathOperator{\diag}{diag}
\DeclareMathOperator{\tr}{tr}

\DeclarePairedDelimiter\abs{\lvert}{\rvert}
\DeclarePairedDelimiter\norm{\lVert}{\rVert}

\DeclarePairedDelimiter\ket{\vert}{\rangle}
\DeclarePairedDelimiter\bra{\langle}{\vert}
\DeclarePairedDelimiterX{\braket}[2]{\langle}{\rangle}{#1\delimsize\vert\mathopen{} #2}
\DeclarePairedDelimiterX{\matele}[3]{\langle}{\rangle}{#1\delimsize\vert\,\mathopen{}#2\,\delimsize\vert\mathopen{}#3}

\usepackage{scalerel,relsize}
\newcommand{\sdagger}{{\mathsmaller{\dagger}}}

\newcommand{\phsdag}{{\phantom{\sdagger}}}

\newcommand{\so}{\mathsmaller{\mathrm{SO}}}
\newcommand{\rB}{\mathrm{B}}
\newcommand{\rC}{\mathrm{C}}
\newcommand{\rL}{\mathrm{L}}
\newcommand{\rR}{\mathrm{R}}
\newcommand{\rZ}{\mathrm{Z}}
\renewcommand{\ss}{\mathrm{s}_\sigma}
\newcommand{\ssb}{\mathrm{s}_{\bar{\sigma}}}

\newcommand{\InF}{\mathrm {InF}}
\def\csqrt{\mathpalette\DHLhksqrt}
\def\DHLhksqrt#1#2{%
\setbox0=\hbox{$#1\sqrt{#2\,}$}\dimen0=\ht0
\advance\dimen0-0.2\ht0
\setbox2=\hbox{\vrule height\ht0 depth -\dimen0}%
{\box0\lower0.4pt\box2}}

\newcommand{\Eqref}[1]{Eq.~\!\eqref{#1}}
\newcommand{\Figref}[1]{Fig.~\!\ref{#1}}

\usepackage{tikz}

\begin{document}
\title{Spin-Orbit Interaction Enabled High-Fidelity Two-Qubit Gates}
\author{Jiaan Qi}\email{qija@baqis.ac.cn}
\author{Zhi-Hai Liu}\email{liuzh@baqis.ac.cn}

\affiliation{Beijing Academy of Quantum Information Sciences, Beijing 100193, China}
\author{H. Q. Xu}\email{hqxu@pku.edu.cn}
\affiliation{Beijing Academy of Quantum Information Sciences, Beijing 100193, China}
\affiliation{Beijing Key Laboratory of Quantum Devices, Key Laboratory for the Physics and Chemistry of Nanodevices, and  School of Electronics, Peking University, Beijing 100871, China
}

\begin{abstract}
We study the implications of spin-orbit interaction (SOI) for two-qubit gates (TQGs) in semiconductor spin qubit platforms. 
SOI renders the exchange interaction governing qubit pairs anisotropic, posing a serious challenge for conventional TQGs derived for the isotropic Heisenberg exchange.  
Starting from microscopic level, we develop a concise computational Hamiltonian that captures the essence of SOI, and use it to derive properties of the rotating-frame time evolutions. Two key findings are made.  First, 
for the controlled-phase/controlled-Z gate, we show and analytically prove the existence of ``SOI nodes'' where the  fidelity can be optimally enhanced, with only slight modifications in terms of gate time and local phase corrections. Second, we discover and discuss novel two-qubit dynamics that are inaccessible without SOI---the reflection gate and the direct controlled-not gate. The relevant conditions and achievable fidelities are studied for the direct controlled-not gate. 
\end{abstract}
\maketitle 

\section{Introduction}
Semiconductor spin qubits are promising candidates for large-scale, fault-tolerant quantum computers due to their potential in industrial scalability and miniaturization in physical size \cite{garciadearquer2021semiconductor}. 
In this scheme,
quantum dots defined in semiconductor devices are commonly used to trap electrons or holes, whose spin states are employed to host quantum information \cite{Loss1998Quantum}. 
Currently, high-speed and precise
manipulation of single-qubit states is dominantly achieved using rf electric signals \cite{vandenBerg2013Fast,Yoneda2018quantumdot,Froning2021Ultrafast,Mills2022HighFidelity}. The underlying principle behind this technique is spin-orbit interaction (SOI), a key mechanism derived from the relativistic Dirac equation and responsible for many novel effects in mesoscopic physics such as the spin Hall effect \cite{Kato2004Observation}, spin transistors \cite{sarma2004spintronics}, and Majorana states in superconductor-coupled nanowires \cite{Sau2010Generic,Oreg2010Helical,Mourik2012Signatures,Deng2012Anomalous}. 

Strong SOI is desirable for enabling  high-speed single-qubit operation \cite{Golovach2006EDSR, Li2013Controlling}. This constitutes one key reason for the recent push for hole spin qubits \cite{Watzinger2018germanium,Froning2021Ultrafast}. 
The role of SOI and its influence in two-qubit gates (TQGs) are, however, less straightforward:
Seminal works on spin-based quantum computing have universally adopted the simplifying assumption that the inter-spin coupling is described by the Heisenberg exchange interaction \cite{Meunier2011Efficient,Russ2018Highfidelity},
\begin{equation*}
 H_\mathrm{ex} = J \,\mb{S}_1 \cdot \mb{S}_2,
\end{equation*}
where $\mb{S}_1$ and  $\mb{S}_2$ are the spin operators for the first and second qubit, with the exchange energy $J$ characterizing the coupling strength. The Heisenberg exchange interaction is spherically symmetric with respect to the reference frame and works well for systems with negligible SOI. 
While it has long been understood in condensed matter theory that SOI gives rise to \emph{anisotropic} exchange coupling, possessing only axial symmetry \cite{Kavokin2001Anisotropic,Kavokin2004Symmetry,Stepanenko2003Spinorbit,Baruffa2010Theory,*Baruffa2010Spinorbit,Li2014Anisotropic}. 
Specially, the exchange coupling is described by a tensor $\mathcal{J}$ that permits the decomposition
\begin{equation}\label{eq:Hex-aniso}
   H_\mathrm{ex} = \mb S_1  \,  \mathcal{J} \, \mb S_2= J_\mathrm{iso} \, \mb S_1  \cdot \mb S_2
  + \mb D\cdot \mb S_1 \times \mb S_2 + \mb S_1
  \, \Gamma\, \mb S_2,
\end{equation}
where $J_\mathrm{iso}$ is the isotropic exchange energy,  $\mb D$ is the so-called  Dzyaloshinskii-Moriya (DM)  vector and $\Gamma$ is a symmetric tensor of rank-1 \cite{Kavokin2004Symmetry,Liu2018Control}. 
The (last two) anisotropic components of exchange coupling in \Eqref{eq:Hex-aniso} is typically small in bulk materials \cite{Kavokin2001Anisotropic}, but could be well-comparable to the isotropic part in low-dimensional semiconductor structures \cite{Kavokin2004Symmetry,Baruffa2010Theory,*Baruffa2010Spinorbit}. 
This discrepancy should not be dismissed lightly, for high-fidelity multiple-qubit gates  are fundamental to  fault-tolerant quantum computing \cite{knill1998resilient,aharonov2008faulttolerant,Xue2022Quantum}.
As manufacturing capability evolves, the prospect of an unknown and uncontrolled fidelity loss  due to SOI is increasingly relevant and demands urgently for a clear understanding  and practical treatment for it.

Much research interest has been drawn to the issue of SOI/anisotropy in recent years. 
One primary perspective considers the deviation to the Heisenberg exchange as an apparent source of error in TQGs.  Various schemes to alleviate the ``anisotropy error'' have been proposed, typically involving tailoring the control pulses and/or the parameters of the exchange interaction \cite{Bonesteel2001Anisotropic,Burkard2002Cancellation,Hao2007Quantum,Zhang2007Interplay,Hao2008Swap,Guerrero2008Effect,Zhou2014swap}.
An alternative perspective is to accept the inevitability of anisotropic exchange and build new sets of TQGs that intrinsically accounts for its effects \cite{Wu2002Universal,Stepanenko2004Universal,Flindt2006SpinOrbit,Milivojevic2017Effective,*Milivojevic2018Symmetric}. 
Both of these treatments require extra resources to account for SOI. These overheads  increase system complexity and may introduce additional control errors by themselves.  
A perhaps more fundamental approach to the issue starts from a microscopic description of the SOI, bypassing the phenomenological expressions like \Eqref{eq:Hex-aniso}.
Such studies combining SOI and quantum computing have been carried out for single-electron driving in double-dot \cite{Khomitsky2012Spin}, spin transport in the presence of SOI  \cite{Li2018Qubit,Zhao2018highfidelity}. Notably,  recent works on microwave-driving controlled-rotation gates on silicon hole spin qubits 
have revealed SOI sweet-spot in suppressing leakage errors while allowing fast operation frequency \cite{Geyer2022Twoqubit,Spethmann2023Highfidelitya}. 

In this paper, we present a simple yet generally applicable solution to the SOI issue in TQGs with direct-current (DC) control \footnote{By DC control, we refer to the time evolution where the Hamiltonian is static in time.}, hopefully clearing the mist of SOI-lead anisotropy. Our paper invites SOI as a key control and optimization parameter in TQGs. This is practically feasible since the SOI coupling strength can be electronically controlled by tunning the gate voltage \cite{Nitta1997Gate} to a suitably large strength \cite{Bosco2021Hole}. 
In Section~\ref{sec:compH}, we derive a concise computational Hamiltonian from microscopic theory and show that it is bidirectionally consistent with the phenomenological description of the anisotropic exchange coupling.  
In Section~\ref{sec:CPhase}, we consider the  controlled-phase (CPhase) gates in the presence of SOI. It is shown that with modified local phase corrections, SOI enables direct high-fidelity implementation of the gate, with optimal fidelity achieved on particular SOI nodes. In Section~\ref{sec:CNOT},  we demonstrate that novel SOI-enabled high-fidelity gates, such as  the reflection gate and  the direct controlled-not (CNOT) gate have become possible under SOI.
Our work  suggest new possibilities by proper utilization of SOI.  and we hope it could be relevant in the future design of spin qubit systems.

\section{The SOI Hamiltonian and anisotropic exchange}\label{sec:compH}
\begin{figure}[htb]
\centering
\includegraphics[width=\linewidth]{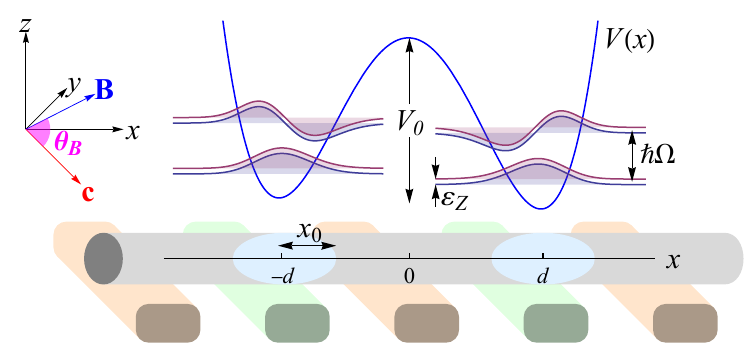}
\caption{Schematic illustration of a model two-spin qubit device defined in, e.g., a semiconductor nanowire with strong SOI via a finger-gate technique. The electrostatic potential profile (in blue) and the first 8 low-lying single-electron states staggered by their energy levels are illustrated on top. 
The top-left inset shows the reference frame and the angle $\theta_{\rB}$ defined between magnetic field $\mb{B}$ and the SOI vector $\mb c$. }
\label{fig:device}
\end{figure}

Our physical model is a spin-qubit quantum device defined in a semiconductor nanostructure with strong SOI. 
The detailed structure of the device could be a nanowire or a two-dimensional dot array. To implement joint evolution of neighboring spin pairs, we isolate on a double-quantum-dot (DQD) subsystem from other spins by applying 
electrostatic confinement, such that electron motion is permit only in the direction along the two dots. 
 For illustrative purpose, we consider a nanowire DQD example in \Figref{fig:device}, where a set of barrier and plunger gates is employed to define a double-dot potential $V(x)$ with minima located at $x=\pm d$. 
 The DQD is maintained in a low-energy, half-filling state (containing two electrons only) and subject to a static and possibly inhomogeneous magnetic field $\mb B (\mb r)$. 
 Some physically achievable constraints are placed to the system. 
First, the dot separation should be large enough $d>x_0$ such that the orbital states are well-localized. Here,  $x_0\equiv\sqrt{\hbar/m_e^\star \Omega}$ is the Bohr radius of the local dot potentials of characteristic frequency $\Omega$ and $m^\star_e$ is the electron effective mass. Second, the characteristic barrier height $V_{0} \gg \hbar \Omega$ should  allow multiple bounded orbits in each dot.  Finally, the orbital energy $\hbar\Omega$ prevails over the Zeeman energy $E_\rZ$ and the detuning $\varepsilon\equiv V(-d)-V(d)$. 

In general, SOI can consist of the Rashba and the Dresselhaus terms. For the quasi-one-dimensional geometry, the coupling Hamiltonian takes on a simple form:
$H_\so = \alpha\, p_x \sigma_c$,
with coupling strength  $\alpha$ and  the Pauli operator  $\sigma_c=\uvec{c}\cdot \bm{\sigma}$ defined along a vector $\uvec{c}$ uniquely determined by SOI.  Presence of SOI modifies many defining aspects of the spin qubit \cite{Li2013Controlling}.
The spin direction is recognized  in the ``spin frame'' $(\uvec{a},\uvec{b},\uvec{c})$, defined by placing the magnetic field $\mb B$ in the upper 
$\uvec{c}{-}\uvec{a}$ plane with angle $\theta_\mr{B}$ to the $\uvec{c}$-axis.
The quantization energy 
reduces from the 
 bare Zeeman energy $E_\rZ \equiv g \mu_\rB \abs{\mb B}$ ($g$ being the Lang\'e-g factor and $\mu_\rB$ the Borh magneton)
 to $\varepsilon_\rZ = f_\so E_\rZ$, for 
$ f_\so\equiv \csqrt{\cos^2\theta_{\rB} + \ee^{-2(x_0/x_\so)^2} \sin^2\theta_{\rB}}$.
 The qubit states also become entangled in spin and orbital parts,
\begin{equation}\label{eq:rotated-states}
\begin{split}
 \ket{\phi_\Uparrow} &\simeq \cos\,(\vartheta/2) \, \ket{0_{+}} \ket{\uparrow}
+ \sin\,(\vartheta/2) \, \ket{0_-} \ket{\downarrow},\\
 \ket{\phi_\Downarrow} &\simeq \cos\,(\vartheta/2) \, \ket{0_{-}} \ket{\downarrow}
- \sin\,(\vartheta/2) \, \ket{0_+} \ket{\downarrow},\\
\end{split}
\end{equation}
where $\ket{0_\pm}\equiv\ee^{\pm\ii \hat{x}/x_\so}  \ket{0}$ are the momentum-displaced orbital ground states, with the spin-orbit length $x_{\so}=\hbar/m_e^\star\alpha$, and the mixing angle 
$\vartheta \equiv \arccos(\cos\theta_{\rB}/f_\so)$.

One can apply the Hund-M\"ulliken theory to derive the low-energy Hamiltonian of the DQD system   \cite{Liu2018Control}. 
Linear combination of states \eqref{eq:rotated-states} at each local dot leads to an energetically-truncated basis  $\ket{\Phi_{\rL\Downarrow}}, \ket{\Phi_{\rL\Uparrow}}, \ket{\Phi_{\rR\Downarrow}},\ket{\Phi_{\rR\Uparrow}}$, where 
the subscript $\rL/\rR$ indicates state localized to the left/right dot.
Using the creation/annihilation operators $a^\sdagger_{j\sigma}/a_{j\sigma}$  for the state 
$\ket{\Phi_{j=\rL,\rR,\sigma=\Uparrow,\Downarrow}}$,
we can write the Hamiltonian as a sum of the dot term
$H_\mr{d} =\sum_{j}   (\varepsilon_j + \varepsilon_{\rZ,j}) n_{j \Uparrow} + (\varepsilon_j - \varepsilon_{\rZ,j}) n_{j \Downarrow}$, 
the tunnelling term 
$H_\mr{t} = \sum_{\sigma} (
 t_{\sigma\sigma} \, a^{\sdagger}_{\rL\sigma} a^{\phsdag} _{\rR\sigma}
 + t_{\sigma\bar\sigma} \, 
a^{\sdagger}_{\rL\sigma} a^{\phsdag}_{\rR\bar\sigma} + h.c.) $ and the Coulomb term
$H_\mathrm{C}= \tfrac{1}{2} U  \,\sum_{j,\sigma} n_{j\sigma} n_{j\bar\sigma}$, 
with $n_{j\sigma}\equiv a_{j,\sigma}^\sdagger a_{j\sigma}$, 
the on-site energy  $\varepsilon_{\rL/\rR}=\pm  \varepsilon/2$,
the charging energy $U$, and   the spin-dependent tunneling coefficients $t_{\sigma\sigma}$ and $t_{\sigma\bar\sigma}$. 
A key insight is that the tunneling coefficients are inter-related in terms of $\vartheta$ and the \emph{relative} SOI strength $\gamma_\so \equiv 2d/x_{\so}$,
\begin{equation}\label{eq:tunneling-coeffs}
\begin{aligned}
t_{\Downarrow\Downarrow} &=t^*_{\Uparrow\Uparrow} = t_0 \bigl( \cos \gamma_\so - \ii \sin\gamma_\so \cos\vartheta\bigr) &&\equiv t,\\
t_{\Downarrow\Uparrow} &= t_{\Uparrow\Downarrow} = t_0\bigl( -\ii \sin\gamma_\so\sin\vartheta\bigr) &&\equiv s.
\end{aligned} 
\end{equation}
Here the spin-conserved ($t$)  and 
spin-flipped  ($s$)  tunneling coefficients are expressed in terms of $t_0=  \sqrt{|t|^2+|s|^2}$, a common factor that only depends on the interdot spacing $d$, potential detuning $\varepsilon$ and the barrier height $V_0$. No particular detail of the potential is otherwise required.

The  quantum computational basis comprises the two-electron states $\{\ket{\uparrow \uparrow},\ket{\uparrow \downarrow},\ket{\downarrow \uparrow},\ket{\downarrow \downarrow}\}$, which are the antisymmetric, half-filling combinations of the single-electron basis states.
Higher-energy two-electron states $\ket{S(2,0)}$ and $\ket{S(0,2)}$ have non-trivial  influence on the $(1,1)$ subspace through virtual tunneling. Their effects can be kept to $O((t_0/U)^4)$ via a transformation.  Combing \Eqref{eq:tunneling-coeffs}, we can eventually deduce the effective Hamiltonian for our  spin-orbit coupled DQD system as,
\begin{equation}\label{eq:Heff}
 H = H_0 - J \ket{\xi}\bra{\xi}.
\end{equation}
Here $H_0 \,\widehat{=}\,  \diag( \varepsilon_\rZ,\tfrac{1}{2}\delta\varepsilon_\rZ,-\tfrac{1}{2}\delta\varepsilon_\rZ,- \varepsilon_\rZ)$ defines the qubit quantization energies, with the average and difference  Zeeman energy $\varepsilon_\rZ \equiv (\varepsilon_{\rZ,\rL}+\varepsilon_{\rZ,\rR})/2$ and $\delta \varepsilon_\rZ \equiv \varepsilon_{\rZ,\rL}-\varepsilon_{\rZ,\rR}$. The coupling part is specified by the exchange energy 
 $J \equiv 2t_0^2(\frac{1}{U-\varepsilon}+\frac{1}{U+\varepsilon})$ and an 
entangled state 
\begin{equation}\label{eq:xi}
  \ket{\xi} \equiv \left(s^* \ket{\uparrow\uparrow} +t^* \ket{\uparrow\downarrow}
  - t \ket{\downarrow\uparrow} + s \ket{\downarrow\downarrow}  \right)/(t_0\sqrt{2}),
\end{equation}
or alternately 
$\ket{\xi} \propto  \ket{\uparrow} \ket{\nearrow}- \ket{\downarrow} \ket{\swarrow}$  by introducing a pair of ``precessed'' states
$\ket{\nearrow}$ and $\ket{\swarrow}$ from the above. Detailed derivation steps to the above are available in Appendix~\ref{app:Hcomp}.

To verify that our computational  Hamiltonian indeed gives rise to the anisotropic exchange coupling, we expand \Eqref{eq:Heff}  under the Pauli basis, $H = \mb S_\rL \cdot  \mb{B}_\rL  + \mb S_\rR\cdot \mb{B}_\rR  +\mb S_\rL  \,  \mathcal{J} \, \mb S_\rR$,
with the local spin operators  $\mb S_{\rL/\rR}=\frac{1}{2}\bm \sigma_{\rL/\rR}$, effective magnetic fields  $\mb{B}_{\rL/\rR}$  and the exchange tensor $\mathcal J$.
In absence of SOI,  $\mathcal J$ is just a scalar, recovering the conventional Heisenberg exchange.
With nontrivial SOI, $\mathcal J$ turns anisotropic.  Performing spherical tensor decomposition, we recover the form in \Eqref{eq:Hex-aniso}, 
with 
\begin{equation}
  \begin{aligned}
    J_0 &= J \,\cos2\gamma_\so ,\\
    \mb{D} &= J\,\sin 2\gamma_\so \,\hat{\mb{v}}, \\ 
    \Gamma &= J \,2\sin^2\!\gamma_\so \,\hat{\mb{v}} \hat{\mb{v}},
  \end{aligned}
\end{equation}
for the vector  $\hat{\mb{v}} = (-\sin \vartheta,0,\cos\vartheta)$.
These expressions are consistent with literatures on anisotropic exchange  (e.g., Ref.~\!\onlinecite{Kavokin2004Symmetry}). 
It is quite remarkable that just a single state $\ket{\xi}$ encodes all information required for the dynamics. On the other hand, compared with \Eqref{eq:Hex-aniso}, the computational Hamiltonian \Eqref{eq:Heff}, combined with the state \Eqref{eq:xi} makes study of TQGs much easier.

\section{SOI-enabled high-fidelity gates}\label{sec:CPhase}
\subsection{Rotating frame and local phase corrections}

Equipped with the computational-space Hamiltonian, 
we are now well-poised to study the time-evolution of qubit pairs. 
But before delving into details, we need to clarify issues regarding to the rotating frame and local phase corrections.
 
Quantum gates are active transformations on quantum states. 
Ideally, a quantum state should remain static when the control Hamiltonian is turned off. 
While an ``intrinsic'' Hamiltonian $H_0$ responsible for the qubit quantization energies generates a common passive rotation $R(\tau)$ of quantum states. 
To offset this effect,  quantum states are typically defined 
in the rotating frame by  applying a reverse rotation to the lab-frame states $ \ket{\psi(\tau) }_\mr{rf}= R^{-1} \ket{\psi(\tau)}_\mr{lab}$ \cite{Russ2018Highfidelity}. Naturally, all quantum gates should be understood in this frame. 
Splitting the total Hamiltonian into $H = H_0+ H_1$, the interaction-picture Hamiltonian
 $\wt H(\tau) = R^{-1}(\tau) H_1 R(\tau)$ is  responsible for generating 
  the  quantum gate through
$\wt{U}(\tau) = \mathcal{T} \exp\left(-\ii \int_0^\tau \wt H(\tau') \dd \tau' \right)$,
  where $\mathcal{T}$ is the time ordering operator. The exact Dyson series is typically difficult to solve. 
Alternatively, we can reversely rotate the lab-frame time evolution operator $U(t)$ and get
\begin{equation}\label{eq:rf-timeevo}
 \wt U(\tau) = R^{-1}(\tau) U(\tau) = \ee^{\ii H_0 \tau} \ee^{-\ii (H_0+H_1)\tau},
\end{equation}
where we have assumed time-independent Hamiltonian within the time frame involved for the matrix exponential expression. In general $\ee^{\ii H_0 \tau} \ee^{-\ii (H_0+H_1)\tau} \neq \ee^{-\ii H_1 \tau}$. Particularly  when $\tau$ is large, both sides are not even close, as the relevant Baker–Campbell–Hausdorff series no longer converges. Dramatic simplification is required for studying properties of $\wt U(\tau)$, which we will show is possible for our effective Hamiltonian  in \Eqref{eq:Heff}.

\begin{figure}[ht]
\centering
\begin{tikzpicture}[x=\linewidth, y=\linewidth]
\node at (0.45,0){ 
\includegraphics[width=0.8\linewidth]{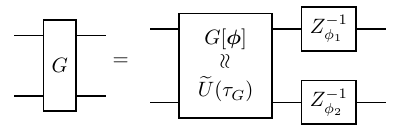}};
\node at (0.,0.1) {(a)};
\node at (0.45,-0.3) { 
\includegraphics[width=0.8\linewidth]{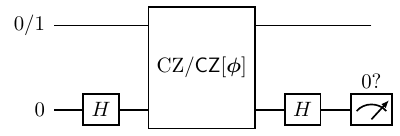}};
\node at (0.,-0.15) {(b)};
\end{tikzpicture}
\caption{(a) The circuit implementation of a generic TQG $G$ with local phase freedom. 
(b) The circuit to determine local phases of an uncorrected $\CZ$ gate.}
\label{fig:localphase}
\end{figure}

The issue of local phase corrections arises mainly from technical perspectives:
It is often difficult to directly realize the target gate from the time evolution.
Instead, experimentalists may allow any gate that differs from the target gate by local phase gates, $Z_{\phi}\equiv \ee^{-\ii (\phi/2) \sigma_Z}$  \cite{Meunier2011Efficient,Veldhorst2015twoqubit}. Moreover, as is with all quantum states, there can be an arbitrary global phase factor as well. 
Taking these phase degrees of freedom into account, we define
\begin{align}\label{eq:CZphi}
G[\bm \phi] \equiv
 \{ \,  \ee^{\ii \phi_0}  \left( Z_{\phi_1}\otimes Z_{\phi_2} \right) G
 \;\vert\;  \forall \phi_0,\phi_1,\phi_2 \}.
\end{align}
Now the target is not a single point  in the $\mathsf{SU}(4)$ space, but rather a three-dimensional manifold charted by the phase parameters $(\phi_0,\phi_1,\phi_2)$. 
By turning on a predetermined coupling Hamiltonian, the rotating-frame evolution operator $\wt U(\tau)$ would come across the target manifold $G[\bm \phi]$ at a certain time $\tau_G$.  This produces the target gate $G$ when combined with local phase corrections  on individual qubits after the transformation, as shown by the circuit model in \Figref{fig:localphase}(a). The justification for introducing local phase gates is that that these single qubit $Z$-axis rotations can be ``virtual'' and need not be actually performed \cite{Vandersypen2004NMR}. One can complete eliminate these local phase gates by permuting them towards the start or the end of the circuit,  or  by shifting the microwave phases for single qubit $X$/$Y$ drives.
Despite the apparent freedom in local phases,  we should point out that  \emph{knowledge} of their values is still important, e.g., at the circuit compilation step, for they can affect final measurement outcomes. Consider for example the circuit in \Figref{fig:localphase}(b), where the upper and lower line represents the control and target qubit and $H$ is the Hadamard gate.
For standard $\CZ$, the circuit outputs state $\ket{00}$ for the input $\ket{00}$ and $\ket{11}$ for input $\ket{10}$. Hence the measurement on the second qubit has full visibility. This is not the case if the $\CZ$ gate is replaced with a $\CZf$ that has not been phase-corrected, where  the possibility for measuring a $\ket{0}$ state on the target qubit would be:
\begin{equation}
 P\left(\text{target}= \ket{0}\right) =
 \begin{cases}
\cos^2(\phi_2/2), & \text{   input} = \ket{00} \\
\sin^2(\phi_2/2), &  \text{   input} = \ket{10} 
 \end{cases}.
\end{equation}
We attribute the reduced visibility of the oscillation signals in Ref.~\onlinecite{Veldhorst2015twoqubit} to this reason, and  this problem  is subsequently addressed in  Ref.~\onlinecite{Watson2018programmable}.
 One the other hand, by placing the Hadamard gates on the first line, $\phi_1$ can also be measured in a similar fashion. Therefore this circuit setup can be used as a protocol to calibrate the local phase corrections if they are not known in advance.

\subsection{The CPhase/CZ gate}
Our first and foremost case study is the CPhase gate, which applies a conditional phase shift to the target qubit. 
A notable member is the controlled-Z (CZ) gate, a universal TQG defined for controlled $\pi$-phase shift.
First proposed in Ref.~\!\onlinecite{Meunier2011Efficient} for systems without SOI, 
the DC implementation of CZ has become the \emph{de facto} way to perform TQGs in state-of-the-art systems \cite{Veldhorst2015twoqubit,Watson2018programmable,Takeda2022Quantum}. Given its importance, here we exclusively focus on the CZ gate and our results can be extended to the general  CPhase gates by simply replacing the $\pi$ phase with an arbitrary phase of interest.

In the computational basis, CZ is represented as the diagonal matrix $\diag(1,1,1,-1)$. Taking account into the local and global phase freedom, we introduce the CZ-class:
\begin{equation}\label{eq:CZ-localPhase}
  \CZf  =\ee^{\ii \phi_0} \diag(1,\ee^{\ii \phi_2},\ee^{\ii \phi_1},
-\ee^{\ii (\phi_1 +\phi_2)}),
\end{equation}
where $\phi_0$ is any global phase, $\phi_{1/2}$ is the phase correction required for the first/second qubit.
An additional conditions that often appear in literatures is the last diagonal term in \Eqref{eq:CZ-localPhase} equating 1, where the phase for the parallel states vanishes and one only need to accumulate a combined phase $\phi_1+\phi_2=\pi$ for antiparallel states. 

To characterize the accuracy of gate implementation, a commonly used indicator is the trace distance function $\tfrac{1}{2}\norm{G-G'}_1$. 
But here we choose to use is the infidelity between gates,
$\InF(G,G')=1-\operatorname{Fid}(G,G')$.
The infidelity function is mutually bounded with the trace distance and permit easier analysis.
For the CZ gate, the instantaneous infidelity function $\InF_\CZ(\tau)$ is define by optimizing the infidelity of the time evolution operator $\wt U(\tau)$ to $\CZf$ with respect to the phase factors. 
\begin{align}\label{eq:InfCZ-def}
\InF_\mathrm{CZ}(\tau) =  \min_{-\pi<\phi_1,\phi_2\le \pi}\frac{d^2 - \abs[\big]{ \tr \bigl(\CZf^\dagger \wt U(\tau) \bigr)}^2 }{d + d^2},
\end{align}
where the Hilbert space dimension $d=4$ here. We have restrict the domain since the infidelity function is invariant with respect to global phase and is period in the phase factors. A further optimization with respect to time identifies the CZ gate infidelity, the gate time, and also simultaneously the optimal phase corrections. 
\begin{equation*}
\begin{aligned}
 &\InF_\mathrm{CZ}\equiv \min_\tau\InF_\mathrm{CZ}(\tau)  , \quad
 \tau_\CZ = \arg\min_\tau \InF_\mathrm{CZ}(\tau)\\
&(\phi_\rL,\phi_\rR) = \arg\min_{\phi_1,\phi_2} \InF_\mathrm{CZ}(\tau_\mathrm{CZ}).
\end{aligned}
\end{equation*}
Here we use $\phi_\rL$ and $\phi_\rR$ to refer to phase corrections on the left and right dot. Additionally, the gate time $\tau_\CZ$ should be restricted to the first  period since the function $\InF_\mathrm{CZ}(\tau)$ is periodic.\begin{figure}[htbp]
 \centering
\begin{tikzpicture}[x=\linewidth,y=\linewidth]
\node (fig-a) at (0.2,0) {\includegraphics[height=0.3\linewidth]{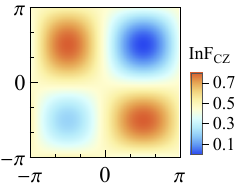}};
\node[rotate=90] (tag-ay) at (-0.01,0.02) {$\phi_2$};
\node (tag-ax) at (0.2-0.02,-0.18) {$\phi_1$};
\node (tag-a) at (-0.01,0.16) {(a)};
\node (tag-b) at (0.44,0.16) {(b)};
\node[rotate=90,scale=0.9] (tag-by) at (0.44,0.02) {$\mathrm{InF}_\mathsf{CZ}(\tau)$};
\node (fig-b) at (0.68,0) {\includegraphics[height=0.3\linewidth]{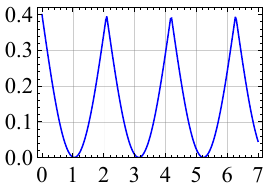}};
\node (tag-bx) at (0.7,-0.18) {$\tau/(\pi/J)$};
\end{tikzpicture}
 \caption{The infidelity of $\wt U(t)$ to the $\CZf$ manifold for a particular energy configuration. 
 (a) The phase-space distribution of the infidelity function at a particular instance $\tau=0.15 \pi/J$. (b) The instantaneous infidelity function $\InF_\CZ(\tau)$. The CZ gate is achieved at the infidelity minima around $\tau=1.05\pi/J$}
 \label{fig:d-phi3}
\end{figure}
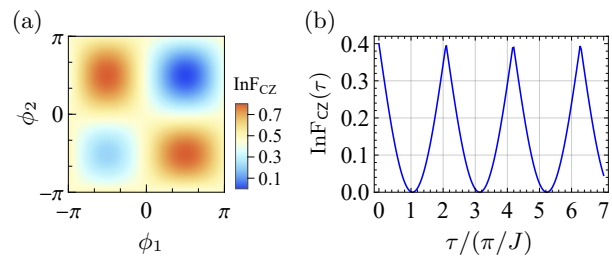
In \Figref{fig:d-phi3}(a), we illustrate the distribution of the gate infidelity within the phase space $(\phi_\rL,\phi_\rR)$ at a particular time instance for an example configuration of $\delta E_\rZ/E_\rZ=0.1$, $J/E_\rZ=0.02$, $\vartheta=0.12\pi$ and $\gamma_\so=0.14\pi$. There can be multiple local minimums in general, and we apply the stochastic differential evolution method to numerically determine the global minimal infidelity at each time instance,  yielding the full $\InF_\CZ(\tau)$ function in \Figref{fig:d-phi3}(b). At $\tau_\CZ\approx 1.05 \pi/J$, $\InF_\CZ(\tau)$ is minimized, and the CZ gate is achieved. 

\begin{figure*}[t]
   \includegraphics[width=\textwidth]{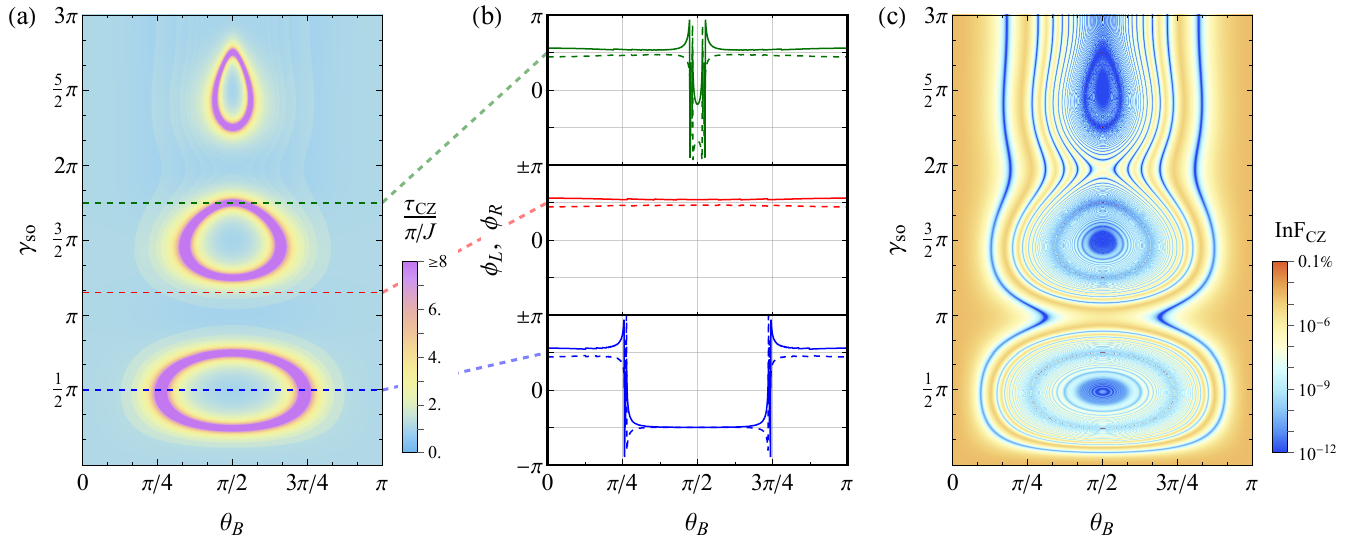}
   \caption{Parameters of the CZ gate under SOI as functions of the magnetic field angle $\theta_{\rB}$ and SOI strength $\gamma_\so$, plotted for a system with $\delta E_\rZ/E_\rZ =0.1$, $J/E_\rZ = 0.02$  and $d/x_0=3$.  (a) Gate time $\tau_\CZ$ in unit of $\pi/J$. The divergent $\tau_\CZ$ values are truncated at $8\pi/J$, which corresponds to the area with oscillatory local phase corrections. (b) Local phases $\phi_\rL$ (solid line) and $\phi_\rR$ (dashed line) plotted along the horizontal line cuts in (a), as indicated by their colors and vertical arrangements. (c) Gate infidelity $\InF_\CZ$ in log scale. The case with no SOI corresponds to the $\gamma_\so =0$ line. 
    }
   \label{fig:CZ}
 \end{figure*}

One can observe through trying out the numerics that provided $\varepsilon_\rZ\gg J$, the $\CZ$ gate is always attainable with high fidelity. But the gate time, local phase corrections and gate infidelity values are subject to the external  control and design parameters $\theta_{\rB}$ and  $\gamma_\so$.
Here, we visualize one set of numerical optimization outcomes in \Figref{fig:CZ} under the 
settings of $\delta E_\rZ/E_\rZ=0.1$, $J/E_\rZ = 0.02$  and $d/x_0=3$.
\Figref{fig:CZ}(a) plots the gate time $\tau_\CZ$, in which one finds a series of  ring-like patterns. The gate time varies slowly around the base value of $\pi/J$ in most cases but diverges within a relatively small window near the rings. 
The finite ring width is due to truncation at $8\pi/J$. The choice is made as an indication that on these truncated areas (rings), the CZ gate cannot be robustly carried out due to rapid oscillatory phase values and divergent evolution time.
 To demonstrate this, we take three  constant $\gamma_\so$ lines and plot $\phi_{\rL/\rR}$  along these lines in \Figref{fig:CZ}(b). In general, the left and right local phase corrections differ by a small amount and vary slowly near an average value of $\pi/2$ when outside and $-\pi/2$ when inside the rings, but change rapidly right on the rings. 
Practically, this should not be too concerning though, as one can circumvent these areas by changing $\theta_{\rB}$ or $\gamma_\so$. Finally,
the gate infidelity $\InF_\CZ$ at $\tau_\CZ$ and optimal $\phi_{\rL/\rR}$ is plotted in \Figref{fig:CZ}(c), showing some quite intriguing oscillating patterns. While the most striking feature is that the infidelity inside the rings turns out to be much \emph{lower} than that outside the rings, indicating that the gate quality could be improved by SOI. 
In particular, one can identify the SOI ``nodes'' at  $\theta_{\rB} = \pi/2$ and 
$\gamma_\so= (2k-1)\pi /2, (k =1,2,\cdots)$, i.e., centers of the divergent rings as
  local minima of the gate infidelity function.

To prove these observation analytically, we consider how the infidelity function can be approximated. According to \Eqref{eq:InfCZ-def}, only the diagonal elements of $\wt U(\tau)$ enters the calculation of gate infidelity since $\CZf$ is diagonal by 
definition. 
Assuming $\varepsilon_\rZ \gg J$, we can determine these diagonal elements with the help of perturbation theory. By assumption, the Hamiltonian $H = H_0 - J \ket{\xi}\bra{\xi}$ can been seen as $H_0$ perturbed by a much smaller term. 
Let us denote the eigenenergies of $H_0$ as $E_n^{(0)} \equiv \matele{n}{H_0}{n}$, where $\ket{n}$ is the  $n$th computational basis state. The perturbed  eigenenergies and eigenstates are denoted as $E_n'= E_n^{(0)} + \delta E_n$ and $\ket{n'}$. 
To calculate the diagonal elements of $\wt U(\tau)$, we invoke \Eqref{eq:rf-timeevo} and take the matrix exponential $\ee^{-\ii H \tau}$ in the eigenstate basis of $H$ and then perform projection onto the diagonal space.
Also since $H_0$ is already diagonal, we have
\begin{equation}
\begin{split}
  &\bra{n}  \wt U(\tau) \ket{n} = \ee^{\ii E_n^{(0)} \tau} \bra{n} \ee^{-\ii H t}\ket{n}  \\
  &= \abs{\braket{n}{n'}}^2 \ee^{-
\ii \delta E_n \tau} +
\sum_{m\neq n}  \abs{\braket{n}{m'}}^2  \ee^{-\ii (E_m'-E_n^{(0)}) \tau },
\end{split}
\end{equation}
where we have split the expression into a summation of a major term ($O(1)$)  originating from the same-energy overlap, and multiple minor terms $O((J/\varepsilon_\rZ)^2)$ originating from correction due to different energy levels.
For our goal of deriving a perturbative expression, it suffice to consider the condition  where the major part implements $\CZf$ and absorb the rest into errors. Comparing with \Eqref{eq:CZ-localPhase}, we see that the DC evolution  reaches the  target $\CZf$ gate when it meet the phase-matching condition:
\begin{equation}\label{eq:phase-matching}
 (\delta E_1  + \delta E_4 -\delta E_2  - \delta  E_3)  \tau = (2k+1) \pi, \quad k \in \mathbb{Z}.
\end{equation}
To derive an appropriate condition, we consider the first order energy perturbations $\delta E_n^{(1)}= - J \abs{ \braket{n}{\xi} }^2$. 
  The phase matches when the evolution time $\tau$ becomes odd multiple of 
\begin{equation}\label{eq:CZ-time}
 \tau_{\CZ} =\frac{\pi/J } {\abs*{\, \abs{\wt t}^2-\abs{\wt s}^2} },
 \end{equation}
 where $\wt t=t/t_0$ and $\wt s = s/t_0$. This expression is still valid after including the second order energy  corrections, which cancel out in \Eqref{eq:phase-matching} due to the symmetric distribution of $H_0$ eigenvalues. 
The divergent condition $\abs{t}=\abs{s}$ produces exactly the rings observed in \Figref{fig:CZ}(a). Physically speaking,  the CZ gate requires a $\pi$-phase difference between the spin-parallel states and the spin-antiparallel states. Yet when $\abs{t}=\abs{s}$, the spin-flipped and spin-conserved tunneling processes are equal in strength, and there is a symmetry in the Hamiltonian between these two sets of states.  A phase difference is unable to accumulate thus the divergent gate time.

After explicitly setting the global and local phase factors to satisfy 
 $\phi_0= -\delta E_1 \tau_\CZ$, 
 $\phi_0+ \phi_\rR=- \delta E_2 \tau_\CZ$ and  $\phi_0+\phi_\rL=- \delta E_3  \tau_\CZ$, we have  the phase of $\diag \wt U_\text{major}$ matches that of $\CZf$.
The corresponding local phase corrections are found to be 
\begin{equation}\label{eq:CZ-local-phases}
 \begin{aligned}
   \phi_\rL &\approx \phi_\rR \approx \frac{\pi}{2}, \quad \text{when } \abs{\wt{t}}>\abs{\wt s}\\
   \phi_\rL &\approx \phi_\rR \approx -\frac{\pi}{2}, \quad \text{when } \abs{\wt{t}}<\abs{\wt s}\\
 \end{aligned}
\end{equation}
at the first order perturbation. Inclusion of the second order perturbation will produces a difference between the two local phase corrections, as detailed in Appendix \ref{app:CZpert}.  When the gate time and the local phase corrections are accurately carried out, the resulting time evolution  is a high-fidelity implementation of the CZ gate. The
gate fidelity  is determined by the non-unity norm of the major part in addition to the minor part. As derived in Appendix \ref{app:CZpert}, we have the infidelity estimation 
\begin{equation}\label{eq:CZ-InF}
  \InF_{\CZ} \lesssim \frac{1}{10}  \Bigl(\frac{J}{\varepsilon_\rZ}\Bigr)^{\!2} \! \Bigg[
\abs{\wt s}^4  +\frac{4\abs{\wt t}^4}{\delta^2} 
+\frac{64(4+\delta^2) }{(4-\delta^2)^2}   \abs{\wt s}^2 \abs{\wt t}^2\Bigg],
\end{equation}
where $\delta \equiv\delta\varepsilon_\rZ/\varepsilon_\rZ$ is a dimensionless number characterizing the Zeeman field gradient.
Our infidelity expression is derived only from the second order eigenenergy and first order eigenstate perturbation, but sufficient to study the qualitative behavior of the full optimization results. 
In particular, the gate-fidelity is not a monotonous function of the SOI strength. An obvious choice for enhancing gate quality is to find a ``sweet spot'' where the RHS of \Eqref{eq:CZ-InF} is kept as small as possible. To find the optimal working condition, we define the square-bracketed terms in the infidelity upper bound as $\kappa$.  Through the relation $\abs{\wt s}^2+\abs{\wt t}^2=1$, we can write $\kappa$ as a function of $\delta>0$ and $x\equiv\abs{\wt s}^2\in [0,1]$:
\begin{equation}
\kappa =x^2+\frac{4 (1-x)^2}{\delta ^2}+\frac{64 \left(\delta ^2+4\right) x (1-x)}{\left(\delta ^2-4\right)^2}.
\end{equation}
After taking the first and second order derivatives with respect to $x$, we find that $\kappa$ attains minimum at either $x=0$ or $x=1$. Combined with the definition 
that $\wt s = -\ii \sin\gamma_\so\sin\vartheta$,
This suggests that the gate infidelity achieves minima either when SOI is absent ($\gamma_\so=0$) or when $\vartheta=\pi/2$ and $\gamma_\so=(k-\tfrac{1}{2})\pi$, namely, at an SOI node. 
Explicitly, for the case without SOI, we have 
\begin{equation}\label{eq:InFCZs0}
  \InF_{\CZ} (\gamma_\so =0) \lesssim \frac{2}{5} \Bigl(\frac{J}{\delta E_\rZ}\Bigr)^2,
\end{equation} 
where we used the bare Zeeman energy $E_\rZ$ here instead of $\varepsilon_\rZ$ since $f_\so=1$.
On the other hand,  at the $k$th SOI node, 
\begin{equation}\label{eq:InFCZkth}
\begin{split}
    \InF_{\CZ} (\gamma_\so&= (k-\tfrac{1}{2})\pi) \lesssim \frac{1}{10} \Bigl(\frac{J}{f_\so E_\rZ}\Bigr)^2\\
   \quad &= \frac{1}{10} \Bigl(\frac{J}{E_\rZ}\Bigr)^2 \exp\!\left[{\frac{(2k \pi-\pi)^2}{8(d/x_0)^2}}\right].
\end{split}
\end{equation} 
It appears as if $f_\so$ is playing the role of magnetic field gradient $\delta$ here.

The implications of \Eqref{eq:InFCZs0} and \Eqref{eq:InFCZkth} are immediate. 
On one hand, to achieve high fidelity gate in system with negligible SOI, a large interdot difference in the qubit energy is required. Thus, in a system with a small $g$-factor variation, device design involving large magnetic field gradient $\delta E_\rZ$ is a theoretical necessity rather than a technical choice. On the other hand, large magnetic field gradient is optional for  quantum dots made in a semiconductor nanostructure with large SOI. 
The fidelity sweet-spot can be attained in systems where the dot geometry matches the SOI strength. In particular, we require 
\begin{equation}
 4d=\pi x_\so,
\end{equation}
for the first SOI node. This condition is within reach for practical systems with $x_\so$ in the order of 100nm \cite{Wang2018Anisotropic}.

\section{SOI-enabled gate dynamics}\label{sec:CNOT}
Looking beyond the CPhase gate, other novel two-qubit dynamics is also enabled after adding the SOI ingredient. Here we briefly discuss two such possibilities---the two-qubit reflection gate and the CNOT gate.

It is well-known that evolving the Heisenberg exchange Hamiltonian for $\tau=\pi/J$ \cite{Loss1998Quantum} produces the SWAP gate, which induces $\ket{\varphi}\ket{\psi}\to\ket{\psi}\ket{\varphi}$ for all single-qubit states $\ket{\varphi}$ and $\ket{\psi}$. 
 Here,  evolving the  anisotropic exchange Hamiltonian $H_\mathrm{ex}=-J\ket{\xi}\bra{\xi}$ leads to 
 \begin{equation}
  \ee^{-\ii (\pi/J) H_\mathrm{ex} } = \mathbb{I} -2 \ket{\xi}\bra{\xi},
 \end{equation}
i.e., a reflection of the two-qubit Hilbert space with respect to 
$\ket{\xi}$.
The reflection gate generalizes over the SWAP gate. The later can be seen as the state reflection with respect to the Bell state $\ket{\Psi^{-}} = \frac{1}{\sqrt{2}} (\ket{\uparrow\downarrow} - \left|\downarrow\uparrow\right\rangle )$.
The reflection gate recovers the SWAP gate for $\gamma_\so=0$, but offers more flexible transformations with finite SOI. 
By adjusting the magnetic field angle and the SOI strength, one can artificially design a large class of state $\ket{\xi}$ that can be used to achieve a single-step reflection.
A high-fidelity implementation of the reflection gate is achievable at the large coupling limit $J\gg \varepsilon_\rZ$---the opposite to the controlled phase gates. 
The gate fidelity can be further improved with optimal local phase corrections.

\begin{figure}[tbp]
 \centering
 \includegraphics*[width=9cm]{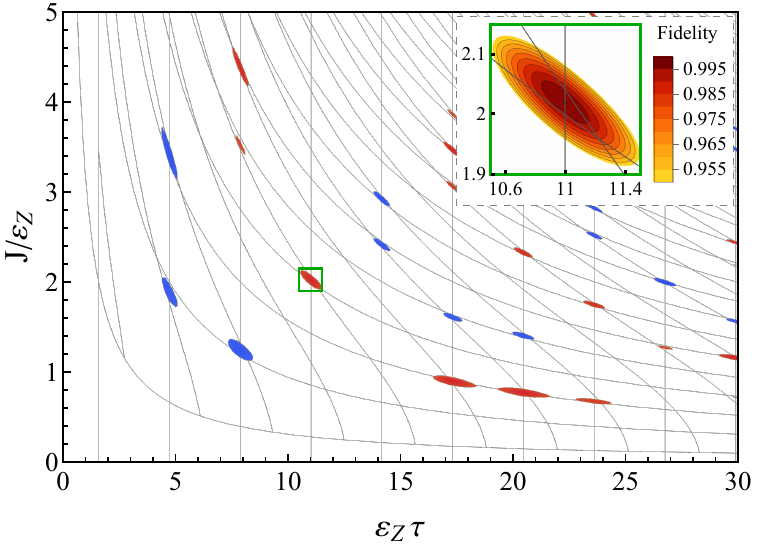}
 \caption{Numerically computed high fidelity ($>95\%$) regions of the two-qubit CNOT gate (blue for $\mathsf{C_2NOT_1}$ and  red for $\mathsf{C_1NOT_2}$) through DC evolution and conditions described in Eq.~\eqref{eq:CNOTcond} (grey lines). The inset shows a zoom-in area where the gate fidelity values are explicitly given. }
 \label{fig:cnot-inf}
\end{figure}

A perhaps more interesting example is the DC implementation of the CNOT gate. Conventionally, the CNOT gate can be implemented in the AC way, i.e., by applying a resonant microwave drive to induce transition between the $\ket{\downarrow\uparrow}$ and $\ket{\downarrow\downarrow}$ states \cite{Zajac2018Resonantly}. Compared with the AC approach, the DC gate implementation, if possible, is more preferred as it uses easier static control signals and is less susceptible to charge noise \cite{Rimbach-Russ2022Simple}.
It would be impossible to attain the CNOT gate by a single-step DC evolution without SOI, as can be proven by showing the commutator 
$[\mathsf{CNOT}, H(\gamma_\so=0)]\neq 0$ for finite exchange coupling $J$. When SOI is present, however, the CNOT gate can be achieved under appropriate conditions. 
Let us use the rotated basis
$\{\ket{\uparrow\nearrow},\{\ket{\uparrow\swarrow},\{\ket{\downarrow\nearrow},\{\ket{\downarrow\swarrow} \}$  for convenience. At the Zeeman splitting condition $E_{\rZ,\rL}=3E_{\rZ,\rR}$,  the Hamiltonian~\!\eqref{eq:Heff} is represented as 
\begin{equation}\label{eq:H-large-detuning}
 H_0-J\ket{\xi}\bra{\xi} \, \mathbin{\widehat{=}} \begin{pmatrix}
 \varepsilon_\rZ & 0 & 0 &0\\
0 & \varepsilon_\rZ  & 0 & 0\\
0 & 0 &  \varepsilon_\rZ  &0\\
0 & 0 & 0 & \varepsilon_\rZ 
 \end{pmatrix} - \begin{pmatrix}
 \frac{J}{2} & 0 & 0 & \frac{J}{2}\\
0 &0 & 0 & 0\\
0 & 0 & 0 &0\\
  \frac{J}{2} & 0 & 0 &\frac{J}{2}
 \end{pmatrix}.
\end{equation}
The representation of $\CNOT$ is in general not diagonal in this basis---unless if the effective SOI strength is fixed at $\gamma_\so=\pi/2$, we have 
\begin{equation}
 \begin{split}
  \mathsf{C_1NOT_2} &\mathbin{\widehat{=}} \diag(1,1,1,-1)\quad  \text{at } \vartheta=\pi/4,\\
\mathsf{C_2NOT_1} &\mathbin{\widehat{=}} \diag(-1,1,1,1)\quad  \text{at }  \vartheta=3\pi/4, 
 \end{split}
\end{equation}
where the subscripts for $\mathsf C$ and $\mathsf{NOT}$ represent the control and target qubit.
In both cases, they transform trivially in the rotating frame defined by $H_0$.  Equating the  interaction picture evolution   $\wt U(\tau)= \ee^{\ii H_0 \tau}\ee^{-\ii H \tau}$ to either CNOT gates, we  obtain three independent
energy-time constraints:
\begin{equation}\label{eq:CNOTcond}
 \begin{aligned}
 & J  \tau=   (2l+1)\pi, \quad 2\varepsilon_\rZ \tau  = (2m+1)\pi,\\
& \csqrt{4\varepsilon_\rZ^2 +J^2} \,\tau= 2 n\pi, \quad \text{where } l,m,n \in \mathbb{N}.\\
\end{aligned}
\end{equation}
If these constraints hold, the CNOT gate can be perfectly achieved (there is no need for local phase corrections).
However, just like the CZ gate cannot be perfectly achieved, these three conditions cannot hold simultaneously, since there is no Pythagorean triple satisfying $(2l+1)^2+(2m+1)^2=(2n)^2$ \cite{Silverman2012Friendly}. But it suffices to achieve a high-fidelity gate provided that these conditions hold \emph{approximately}.
In \Figref{fig:cnot-inf}, we numerically compute the gate fidelity of the $\mathsf{C_1NOT_2}$ and $\mathsf{C_2NOT_1}$ as a function of the dimensionless exchange energy $J/\varepsilon_\rZ$ and evolution time $\epsilon_\rZ \tau$ (in units determined by the qubit energy $\varepsilon_\rZ$). One can see that the high-fidelity regions appear precisely at the approximate intersections of  the constraints in Eqs.~\!\eqref{eq:CNOTcond}. 
Through a suitable combination of the effective SOI strength $\gamma_\so$, the magnetic field  angle $\theta_{\rB}$, the 
exchange energy $J$ and evolution time $\tau$, one can implement the CNOT gate with fidelity surpassing 99.5\% by just a single-step evolution. In Table~\ref{tab:CNOTsol}, we compute and summarize a list of high-fidelity points of the CNOT gate for the range considered in \Figref{fig:cnot-inf}.
This could be used for reference in future experimental implementations.
\begin{table}[htbp]
 \centering
 \begin{tabularx}{9cm}{|>{\centering\arraybackslash}X |>{\centering\arraybackslash}X |>{\centering\arraybackslash}X |>{\centering\arraybackslash}X|}
 \hline
Gate & $\varepsilon_\rZ \tau$ & $J/\varepsilon_\rZ$ & Fidelity \\\hline\hline
$\mathsf{C_2NOT_1}$ & 20.4204 &	4.15737	& 0.99958\\\hline
$\mathsf{C_1NOT_2}$ & 17.2788 &	3.46028	& 0.99918\\\hline
$\mathsf{C_2NOT_1}$ & 26.7035 &	1.99654	& 0.99914\\\hline
$\mathsf{C_1NOT_2}$ & 23.5619 &	4.94016	& 0.99795\\\hline
$\mathsf{C_1NOT_2}$ & 23.5619 &	4.65948	& 0.99772\\\hline
$\mathsf{C_1NOT_2}$ & 7.85398 &	4.37725	& 0.99744\\\hline
$\mathsf{C_1NOT_2}$ & 26.7035 &	3.18416	& 0.99643\\\hline
$\mathsf{C_1NOT_2}$ & 20.4204 &	0.77555 & 0.99599\\\hline
$\mathsf{C_1NOT_2}$ & 10.9956 &	2.02033	& 0.99504\\\hline
$\mathsf{C_1NOT_2}$ & 17.2788 &	0.89983	& 0.99493\\\hline
 \end{tabularx}
 \caption{A selected list of high-fidelity solutions of \Eqref{eq:CNOTcond} for the CNOT gate, sorted by the estimated optimal fidelity. }
 \label{tab:CNOTsol}
\end{table}

\section{Conclusions}
In summary, we have studied the role of SOI in the dynamics of a spin qubit pair. For an exchange-coupled two-electron spin system, the compact computational Hamiltonian of \Eqref{eq:Heff} is derived and the connection of SOI to the anisotropic exchange interaction is revealed. Using this Hamiltonian, we deliver the central message that SOI needs not be treated as a noise source in TQGs, but rather can be naturally taken as advantages. 
Specifically, it is shown that the CPhase gate can be accurately implemented with adaption of SOI, and the gate fidelity is optimized at certain practically attainable SOI nodes. It is also shown that  SOI enable two-qubit dynamics that are conventionally impossible, with the general  reflection gate and the CNOT gate examples explicitly constructed. 
Finally, we must point out that the dynamics considered here is purely unitary and  errors involved are coherent in nature. This simple model should serve to demonstrate the power of SOI for facilitating high-fidelity gates and for creating novel possibilities in constructing spin-based quantum computing chips.

\begin{acknowledgments}
This work is supported by the Beijing Postdoctoral Research Foundation 
(Grant No.~\!2023-zz-050), the National Natural Science Foundation of China (Grant Nos.\ 92165208 and 11874071) and the Key-Area Research and Development Program of Guangdong Province (Grant No.\ 2020B0303060001). We would like to thank Ji-Yin Wang and Yi Luo for helpful discussions.
\end{acknowledgments}

\bibliographystyle{apsrev4-2}
\bibliography{SOITQG_bib}

\newpage
\onecolumngrid
\appendix

\section{Deriving the computational Hamiltonian}\label{app:Hcomp}
\subsection{Frame setup and problem outline}
In general, we can write the total Hamiltonian as a sum of the single-electron Hamiltonian and electron-electron interaction Hamiltonian, 
\begin{equation}\label{eq:Htot}
    H = \sum_{i=1,2} H_{e}(\mb r_i) + 
\frac{e^2}{4\pi\varepsilon_\mr{e}\abs{\mb r_1 - \mb r_2}},
\end{equation}
where $\varepsilon_\mr{e}$ is the electric permittivity of the device.
$H_e$ consists of 
\begin{equation}\label{eq:Hse}
     H_{e}(\mb r)  = 
     \frac{\mb p^2}{2m_e^\star}  +V(\mb r) + H_\so + H_\rZ,
\end{equation} 
where $H_\so$ is the SOI Hamiltonian and $H_\rZ$ is the Zeeman Hamiltonian.  The momentum operator is given by $\mb p = -\ii \hbar \nabla + e \mb A$ under the magnetic field $\mb B(\mb r) = \nabla \times \mb A$. We leave out the optional rf electric field term here as it is irrelevant for our DC TQG implementations.

The SOI in a semiconductor can in general consists of the Rashba  and Dresselhaus terms as a result of bulk inversion asymmetry and structural inversion asymmetry. The forms of interaction are normally defined in the crystallographic frame. By choosing $\uvec{x}'\parallel [100]$, $\uvec{y}'\parallel [010]$, $\uvec{z}'\parallel [001]$, and  assuming that the device is fabricated in the (001) plane, we have,
$ H_\so = \alpha_\mathrm{R} \left(p_{x'} \sigma_{y'}-p_{y'}\sigma_{x'} \right)+ \alpha_\mathrm{D} \left(p_{x'} \sigma_{x'}-p_{y'}\sigma_{y'} \right)$,
where $\alpha_\mr{R}$ and $\alpha_\mr{D}$ are the coupling strengths for the Rashba  and Dresselhaus effects respectively. For our DQD system, the electron motion is further restricted to quasi-one-dimension along the dots. 
This defines the ``device frame'' $(\uvec{x}, \uvec{y}, \uvec{z})$, where the quantum dots are lining along the $\uvec{x}$ axis and $\uvec{z}\parallel\uvec{z'}$. Denoting the angle between $\uvec{x}$ and $\uvec{x'}$ as $\varphi_{xx'}$, we can derive, for our DQD system,
\begin{align}
  H_\so &= p_x  \left( \alpha_\mr{R} \sigma_y + \alpha_\mr{D} \cos2\varphi_{xx'}\, \sigma_x - \alpha_\mr{D}\sin2\varphi_{xx'} \sigma_y \right) \notag \\
  &=p_x \,   ( \alpha\, \uvec{c}\cdot \bm\sigma) =  \alpha\, p_x \sigma_c,
\end{align}
where $\uvec{c}\equiv\alpha^{-1} (\alpha_\mr{D}\cos 2\varphi_{xx'}\,  \uvec{x}+(\alpha_\mr{R}-\alpha_\mr{D}\sin 2\varphi_{xx'})\uvec{y} )$ is a unit vector representing the direction of SOI and $\alpha$ is the normalized SOI coupling coefficient. The restriction to one-dimension also suppresses cyclotron movement of electrons and we can neglect the gauge field $\mb A$ as well.

Despite restriction in electron orbital motion, the electron spin can still point to all directions. For systems with strong SOI, spin is more conveniently represented in the ``spin frame''---$(\uvec{a}, \uvec{b}, \uvec{c})$---with spin-up/down along the $\pm \uvec{c}$ direction. To eliminate the remaining uncertainty in $\uvec{a}$ and $\uvec{b}$, we demand that the magnetic field $\mb{B}$ lying in the upper $\uvec{c}{-}\uvec{a}$ plane, such that  $\theta_{\rB}\in[0,\pi]$ denotes the angle between $\mb B$ and $\uvec{c}$. Naturally, $\uvec{b}=\uvec{c}\times\uvec{a}$, and we can identify 
\begin{equation}\label{eq:Hz-pos}
 H_\rZ=\frac{1}{2}E_\rZ(\mb r) \left(\cos\theta_{\rB}\, \sigma_c + \sin\theta_{\rB}\, \sigma_a\right),
\end{equation}
where $E_\rZ \equiv g(\mb r) \mu_\rB \abs{\mb B(\mb r)}$ is the Zeeman splitting energy,  with $g(\mb r)$ being the position-dependent Lang\'e-g factor and $\mu_\rB$ the Borh magneton.

The positional basis used in \Eqref{eq:Htot} to \Eqref{eq:Hz-pos} is inconvenient for studying the low energy dynamics of  the system since the representation is uncountably-infinite dimensional. A much better choice is the single-electron energy eigenstates, where $H_e$ is diagonal and  we can truncate the basis to the first few terms to write down the low energy Hamiltonian.
This scheme is all good except it is hard to find the exact eigenstates. Instead, another orthonormal set of  electron wave function $\{\ket{\Phi_{\mu}}\}$ can be used.  Notice $\ket{\Phi_\mu}$ need not be a product state of the orbital and the spin wave functions. As long as the low energy eigenstates resides in the span of the first few terms of the basis, we may truncate the basis to obtain a matrix of a finite dimension.
Hence the goal now is to find an appropriate basis to support the low energy wave functions. For our DQD system,  we can construct a basis by the ``linear combination  of atomic orbits'' method by first solving for the wave functions for two single dots and then linearly combining the two local dot wave functions to form a DQD basis.

\subsection{The single-dot problem}
Suppose we have a single dot subject to harmonic local potential at the center of the coordinate system. Its static Hamiltonian can be written under the effective mass approximation as 
\begin{equation}\label{eq:Hsd}
 H_\mathrm{sd} = \frac{p^2}{2 m_e^\star} + \frac{1}{2}m_e^\star \Omega^2 x^2+ H_\so + H_\rZ 
\end{equation}
Even for such simple potential profile, analytical solutions cannot be found  in a closed form. 
Writing down its eigenenergies and eigenstates would still require perturbation theory.
Formal treatment of such Hamiltonian can be found in, e.g., Ref.\!~\onlinecite{Golovach2006EDSR} and Ref.\!~\onlinecite{Li2013Controlling}.
We will follow the latter approach here, which uses the Zeeman term as  perturbation, allowing treatment of strong SOI.

First, we define the unperturbed Hamiltonian $H_0$ as the $H_\mathrm{sd}$ in \Eqref{eq:Hsd} excluding $H_\rZ$.
Conjugating $H_0$  with $\ee^{\ii (x/x_\so)  \sigma_c}$ yields, up to a constant, the simple harmonic oscillator Hamiltonian.
Therefore the $H_0$ eigenvalues and eigenstates  are found to be 
\begin{align}
E^{(0)}_{n\sigma} &= \hbar\Omega(n+\frac{1}{2}) -\frac{1}{2}m^\star_e \alpha^2,\\
    \ket{\phi^{(0)}_{n\sigma}} &= \ee^{-\ii \ss x/x_\so} 
   \ket{n} \ket{\sigma}\; \mbox{\rm with}\; \sigma \in\{\uparrow,\downarrow\},
\end{align}
where 
$\ket{n}$ is the $n$th harmonic oscillator eigenstate, and 
$ \mathrm{s}_{\uparrow\downarrow} = \pm1$
is the spin-sign function. 
It is seen that for $H_0$, states with same orbital number $n$ and opposite spins are degenerate. This degeneracy is lifted by turning on $H_\rZ$. 
The degenerate perturbation theory requires finding an appropriate  basis that  diagonalizes $H_\rZ$. For the orbital ground states  in particular, 
\begin{equation}\label{eq:HZn=0}
\bra{0} H_\rZ \ket{0} =\frac{E_\rZ}{2}
 \begin{pmatrix}
 \cos\theta_{\rB} &\ee^{-(\frac{x_0}{x_\so})^2} \sin\theta_{\rB}  \\
 \ee^{-(\frac{x_0}{x_\so})^2} \sin\theta_{\rB}  & -\cos\theta_{\rB} 
 \end{pmatrix}.
\end{equation}
Diagonalizing this matrix, we find the first-order energy corrections and 
the suitably oriented eigenstates to be
\begin{align}
E_{0\sigma}^{(1)}&= \mr{s}_\sigma \tfrac{1}{2} f_\so\, E_\rZ \equiv\ss \tfrac{1}{2}  \varepsilon_\rZ ,\\
\ket{\phi^{(0')}_{0\sigma}} &=\cos(\tfrac{\vartheta}{2}) \,\ket{\phi^{(0)}_{0\sigma}}+\mr s_\sigma \sin(\tfrac{\vartheta}{2}) \, \ket{\phi^{(0)}_{0\bar\sigma}}\label{eq:psi0p},
\end{align}
where we have defined the SOI-related factors
\begin{align}
 f_\so &\equiv \sqrt{\cos^2\theta_{\rB} + \ee^{-2\eta^2_\so} \sin^2\theta_{\rB}},\\
  \vartheta &\equiv \arccos(\cos\theta_{\rB}/f_\so), 
\end{align}
which account for the modifications of the Zeeman energy splitting and the effective magnetic field angle due to the inclusion of SOI. 
For negligible SOI $\eta_\so\ll1$,  $f_\so \approx 1$ and $\vartheta\approx\theta_{\rB}$  for all $\mb B$ field angles $\theta_{\rB}$. in the very strong SOI regime, $\eta_\so\ge2$ and $f_\so \approx \cos(\theta_{\rB})$, and thus the effective $\mb B$ field angle is strongly regulated as $\vartheta\approx 0$ for 
$\theta_{\rB}<\pi/2$ and $\vartheta\approx \pi$  for $\theta_{\rB}>\pi/2$, i.e., either parallel or antiparallel to the SOI vector. For an intermediate SOI strength, the behavior is between the above two limiting cases.

Further application of the non-degenerate perturbation theory  in the rotated basis yields corrections from higher orbital states. Keeping terms only up to the first order in $\xi_Z \equiv  E_\rZ/\hbar\Omega\ll 1$, we have
\begin{equation}\label{eq:psi0perturb}
\ket{\phi^{(1)}_{0\sigma}} =   \ket{\phi^{(0p)}_{0\sigma}}+   \sum_{n=1}^\infty  \chi_{n} \Bigl[ 
\ssb^n \cos(\tfrac{\vartheta}{2}) \,\ket{\phi^{(0)}_{n\bar\sigma}} +  \ss^{n+1}\sin(\tfrac{\vartheta}{2}) \, \ket {\phi^{(0)}_{n\sigma}} \Bigr].
\end{equation}
We have left out the normalization factor here for brevity. The mixing amplitudes with higher orbital states are given by 
$\chi_{n}=   - \tfrac{1}{2}\xi_Z \ee^{-(x_0/x_\so)^2}\sin(\theta_{\rB})   {(\sqrt{2} \ii x_0/x_\so)^n}/{(n\sqrt{n!})}$, which are suppressed gradually as the orbital level $n$ increases.
Similar calculations can be carried out for higher orbital levels as well. This eventually gives rise to a complete basis of states, each with entangled spin and orbital parts.
In this basis,  \Eqref{eq:Hsd} is diagonal up to first order in  $\xi_\rZ$, and its $n=0$ subspace defines the qubit Hamiltonian $H_Q=\frac{1}{2}\varepsilon_\rZ \sigma_c$.
But the electric dipole term 
$\mb E(t) \cdot \mb r$ will not be diagonal  due to nonzero matrix elements between neighboring orbital levels.  
This is the basis for the spin-orbit qubit and single qubit manipulations.

\subsection{The double-dot basis}
The only thing left unspecified in the Hamiltonian \Eqref{eq:Hse} is the electro-static potential $V(\mb r) = V(x)$ of the DQD system. 
In the vicinity of the local minima $x=\pm d$, $V(x)$ is approximately harmonic.
We shift the energy reference level such that $V(\mp d)=\pm \varepsilon/2$ and approximate the local harmonic potentials at $x=-d$ and $x=d$ by $V_\rL$ and $V_\rR$. 
In general, the local harmonic frequencies could differ, but introducing different oscillator for each dot would be overcomplicated if only few lowest orbital levels are of interest. Here we  bypass this issue by assuming that the dot frequencies are of a characteristic magnitude  $\Omega$. $V_{\rL}$ and  $V_{\rR}$ are defined  by
 translating the common potential $V_\rC=\frac{1}{2}m_e^\star\Omega^2x^2$ to $V_{\rL/\rR}=\frac{1}{2}m_e^\star\Omega^2(x\pm d)^2$ and shifting the energy by $\pm \varepsilon/2$.
The differences from the exact local Taylor expansions are absorbed by $\Delta V_L$ and $\Delta V_R$, respectively, such that the total DQD potential can be split as
\begin{equation}\label{eq:local-pot}
 V = V_\rL +\Delta V_\rL =V_\rR +\Delta V_\rR.
\end{equation}

Assuming that the effective interdot barrier is much greater compared to the orbital spacing and the dots are well separated,
the low energy eigenstates are  mostly concentrated at the two single dots localied  around $x=\pm d$.
To capture those localized states, we consider translating a basis $\{\ket{\phi^{\rC}_{n\sigma}}\}$ for the $V_\rC$ potential to the left and right dots,
\begin{equation}\label{eq:psiC-translation}
  \ket{\phi^{\rL}_{n\sigma}} = \ee^{\ii p d}\,  \ket{\phi^\rC_{n\sigma}}, \quad 
    \ket{\phi^{\rR}_{n\sigma}} = \ee^{- \ii pd}\,  \ket{\phi^\rC_{n\sigma}}.
\end{equation}
Further assuming that $\hbar \Omega \gg  \varepsilon$ or  $E_\rZ$,
we can define a set of basis states $O$, ordered by increasing energy,
by interleaving the two sets of basis states in \Eqref{eq:psiC-translation}.
$O$ is over-complete and orthogonalization is required to make proper basis. 
Formulating a systematic orthogonalization procedure is not the focus of this work. 
As a low energy theory, we can truncate the basis to only consider the 
 four lowest orbital states $O_{n=0}$. Given this basis does not involve higher orbital states anyway, it make little sense to use the fully perturbed states in \Eqref{eq:psi0perturb} either. But the degenerate perturbation required for breaking the Kramer degeneracy is necessary as it does not involve higher orbital states. We thus take the  $\xi_Z\ll 1$ limit and consider the properly rotated ground orbital states at the center using \Eqref{eq:psi0p},
\begin{equation}\label{eq:psi-sopair}
 \begin{aligned}
 \ket{\phi^{\rC}_{0\Uparrow}} &=\cos\tfrac{\vartheta}{2}\, \ee^{-\ii x/x_\so} 
   \ket{0} \ket{\uparrow} + \sin\tfrac{\vartheta}{2}\, \ee^{+\ii x/x_\so} 
   \ket{0} \ket{\downarrow} ,\\
    \ket{\phi^{\rC}_{0\Downarrow }} &=\cos\tfrac{\vartheta}{2} \,\ee^{+\ii  x/x_\so} 
   \ket{0} \ket{\downarrow} - \sin\tfrac{\vartheta}{2} \, \ee^{-\ii  x/x_\so} 
   \ket{0} \ket{\uparrow},
 \end{aligned}
\end{equation}
where we follow the double-arrow convention to indicate entanglement in spin and orbital components, as opposed to single arrows for spin only.

To derive an orthonormal low-energy basis for the DQD system, we 
substitute \Eqref{eq:psi-sopair} into \Eqref{eq:psiC-translation} and calculate the state overlaps.
States within the same dot are automatically orthonormal, since they are eigenstates of the local Hamiltonian, 
$\braket{\phi^{\rL}_{\sigma}}{\phi^{\rL}_{\sigma'}}= \braket{\phi^{\rR}_{\sigma}}{\phi^{\rR}_{\sigma'}}=\delta_{\sigma\sigma'}$.
To calculate the interdot overlaps, we apply the vacuum expectation formula of displacement operators and find two distinct types of coefficients, i.e., the state overlaps between same spins and that between opposite spins,
 \begin{align}\label{eq:wavefun-overlaps}
s_d &\equiv \braket{\phi^{\rL}_{ 0\Downarrow}}{\phi^{\rR}_{ 0\Downarrow}} = \braket{\phi^{\rL}_{ 0\Uparrow}}{\phi^{\rR}_{ 0\Uparrow}}^* = \ee^{-({d}/{x_0})^2} \left( \cos\gamma_{\so}-\ii  \sin \gamma_{\so} \cos\vartheta\right),\\[1ex]
s_x & \equiv \braket{\phi^{\rL}_{ 0\Downarrow}}{\phi^{\rR}_{ 0 \Uparrow}}=\braket{\phi^{\rL}_{ 0\Uparrow}}{\psi^{\rR}_{ 0 \Downarrow}} \phantom{^*} =\ee^{-({d}/{x_0})^2} \left( -\ii  \sin\gamma_{\so}\sin\vartheta\right).
\end{align}

To form an orthonormal DQD basis out of the local states, we label the four DQD basis states as $\ket{\Phi_{\rL\Downarrow}}$, $\ket{\Phi_{\rL\Uparrow}}$, $\ket{\Phi_{\rR\Downarrow}}$ and $\ket{\Phi_{\rR\Uparrow}}$, which are linear superpositions of the four local states,
\begin{equation}\label{eq:normalizedstates}
\begin{aligned}
\ket{\Phi_{\rL\Downarrow}}&= 
c_{11} \ket{\phi^{\rL}_{0\Downarrow}} 
+ c_{12} \ket{\phi^{\rL}_{0\Uparrow}}  +c_{13} \ket{\phi^{\rR}_{0\Downarrow}}
    +c_{14} \ket{\phi^{\rR}_{0\Uparrow}},\\
\ket{\Phi_{\rL\Uparrow}}&= c_{21} \ket{\phi^{\rL}_{0\Downarrow}} 
+ c_{22}\ket{\phi^{\rL}_{0\Uparrow}}  +c_{23} \ket{\phi^{\rR}_{0\Downarrow}}
 +c_{24} \ket{\phi^{\rR}_{0\Uparrow}},\\
\ket{\Phi_{\rR\Downarrow}}&= c_{31} \ket{\phi^{\rL}_{0\Downarrow}} 
+ c_{32} \ket{\phi^{\rL}_{0\Uparrow}}  + c_{33} \ket{\phi^{\rR}_{0\Downarrow}}
 +c_{34} \ket{\phi^{\rR}_{0\Uparrow}},\\
\ket{\Phi_{\rR\Uparrow}}&= c_{41} \ket{\phi^{\rL}_{0\Downarrow}} 
+ c_{42} \ket{\phi^{\rL}_{0\Uparrow}}  +c_{43} \ket{\phi^{\rR}_{0\Downarrow}}
 +c_{44}  \ket{\phi^{\rR}_{0\Uparrow}}.
 \end{aligned}
\end{equation}
Applying the orthonormal condition for the basis states leads to a set of 10 independent equations. But there are in total 16 complex coefficients to solve. This indicates that there is a lot of freedom in choosing the basis states. 
For a representation that accurately depicting the low energy dynamics of the DQD system, we impose additional constrains out of locality and symmetry: The DQD basis states should capture the electron wave functions localized at a particular dot with a particular spin and should be symmetric with respect to the dot location and the spin direction.
According to the locality principle, the DQD basis states should be considered as perturbations to the correspondingly  local states. Here we choose the smallness factor to be $\ee^{-({d}/{x_0})^2}$. The premise that 
$d/x_0$ is large is in fact mandatory for well localized dot states.  As the normalization factors can be added afterwards, we demand that $c_{ii}=1$ with $c_{ij}=O( \ee^{-({d}/{x_0})^2})$ for $i\neq j$.  Orthogonality yields a set of 6 independent complex constraints,
\begin{equation}
 \sum_{i,j} c^*_{mi} c_{nj} \braket{\phi_{i}}{\phi_{j}} =0, \quad \text{for}\quad n\neq m,
\end{equation}
where we have used the shorthands $\phi_1,\phi_2,\phi_3$, and$\phi_4$ for 
$\phi^{\rL}_{0\Downarrow}, \phi^{\rL}_{0\Uparrow}, \phi^{\rR}_{0\Downarrow}$, and $\phi^{\rR}_{0\Uparrow}$, respectively.
According to the symmetry principle, transposing the dot or the spin labels should leave the amplitude of the respective coefficients invariant. This produces 9 real constrains,
\begin{equation}
 \begin{aligned}
  &\abs{c_{12}} = \abs{c_{21}} =\abs{c_{34}}=\abs{c_{43}},\\
  &\abs{c_{13}} = \abs{c_{31}} =\abs{c_{24}}=\abs{c_{42}},\\
  &\abs{c_{14}} = \abs{c_{41}} =\abs{c_{32}}=\abs{c_{23}}.
 \end{aligned}
\end{equation}
Still out of symmetry consideration, we further demand the normalization factors to be identical, leading to 3 additional real constraints,
\begin{equation}
\begin{aligned}
 & \sum_{i,j} c^*_{1i} c_{1j} \braket{\phi_{i}}{\phi_{j}} = 
 \sum_{i,j} c^*_{2i} c_{2j} \braket{\phi_{i}}{\phi_{j}}   \\
  =& \sum_{i,j} c^*_{3i} c_{3j} \braket{\phi_{i}}{\phi_{j}}= 
  \sum_{i,j} c^*_{4i} c_{4j} \braket{\phi_{i}}{\phi_{j}} .
\end{aligned}
\end{equation}
We now have 24 real constrains for 24 real unknowns. Therefore, these coefficients can be in principle fully solved, producing a unique DQD basis that is both localized and symmetric. The exact coefficients, however, are quite involved functions of $s_d$ and $s_x$. For our treatments, it suffices to keep these expressions to the leading order in $\ee^{-({d}/{x_0})^2}$. Therefore, we make the shift $s_d\to \epsilon s_d$, $s_x\to \epsilon s_x$ and 
$ c_{ij}\to \epsilon\, c_{ij} + \epsilon^2 d_{ij}+\cdots,\ i\neq j$.
Solving the constraints to the leading order, we have 
\begin{equation}
\begin{aligned}
 c_{12}&=c_{21}=c_{34}=c_{43}=0, \\
 c_{13}&=c^*_{31}=c^*_{24}=c_{42}= -s_d^*/2,\\
  c_{14}&=c^*_{41}=c_{23}=c^*_{32}= -s_x^*/2.
\end{aligned}
\end{equation}
These solutions are in fact good enough for all constraints to remain valid up to $O(\epsilon^2)$, 
Therefore we may write our DQD basis states as 
\begin{equation}\label{eq:DQD-basis}
\begin{aligned}
\ket{\Phi_{\rL\Downarrow}}&= 
\ket{\phi^{\rL}_{0\Downarrow}} 
-\tfrac{1}{2} s_d^* \ket{\phi^{\rR}_{0\Downarrow}}
   -\tfrac{1}{2} s_x^* \ket{\phi^{\rR}_{0\Uparrow}},\\
\ket{\Phi_{\rL\Uparrow}}&= \ket{\phi^{\rL}_{0\Uparrow}} 
 -\tfrac{1}{2} s_d \ket{\phi^{\rR}_{0\Uparrow}}
 -\tfrac{1}{2} s_x^* \ket{\phi^{\rR}_{0\Downarrow}},\\
\ket{\Phi_{\rR\Downarrow}}&= \ket{\phi^{\rR}_{0\Downarrow}}
-\tfrac{1}{2} s_d \ket{\phi^{\rL}_{0\Downarrow}} 
-\tfrac{1}{2} s_x \ket{\phi^{\rL}_{0\Uparrow}},\\
\ket{\Phi_{\rR\Uparrow}}&=\ket{\phi^{\rR}_{0\Uparrow}}
-\tfrac{1}{2} s_d^* \ket{\phi^{\rL}_{0\Uparrow}}
-\tfrac{1}{2} s_x \ket{\phi^{\rL}_{0\Downarrow}} .
 \end{aligned}
\end{equation}

\subsection{Low energy Hamiltonian and tunnelling coefficients}
\begin{figure}[htbp]
 \centering
 \includegraphics[width=8cm]{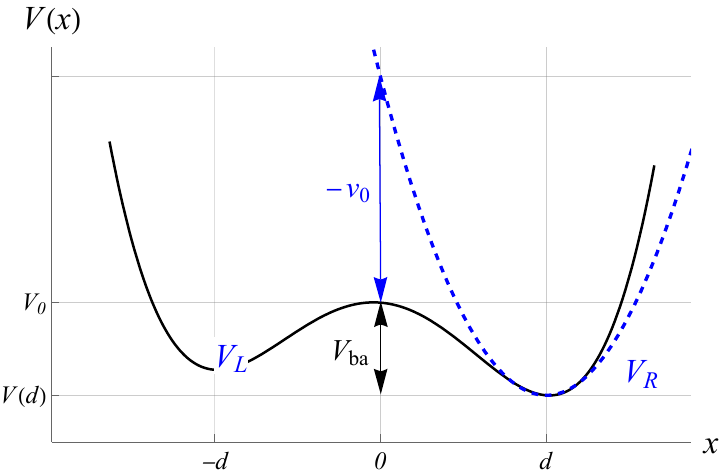}
 \caption{The $v_0$ term as determined by the potential profile}
 \label{fig:DQD-pot-v0}
\end{figure}

Equipped with the low energy basis, we proceed to derive the low energy Hamiltonian for the DQD system. 
In the field operator language, we may represent single-particle and two-particle interaction Hamiltonian by the creation $a^\dagger_{\mu}$ and annihilation operator $a_{\mu}$ for $\ket{\Phi_{\mu}}$ as
\begin{equation}\label{eq:Hamiltonian-fieldOp}
 H_e =\sum_{\mu\nu}  {T_{\mu\nu}} a_{\mu}^\dagger  a_{\nu}, \quad 
 H_{ee} =   \frac{1}{2} \sum_{\mu\nu\rho\lambda} V_{\mu\nu\rho\lambda}  a_{\mu}^\dagger  a_{\nu}^\dagger   a_{\rho}  a_{\lambda}.
\end{equation}
where  the state indices in \Eqref{eq:Hamiltonian-fieldOp} now include both the site and spin-orbit quantum numbers. The coefficients for the $H_e$ are given by 
$T_{X\sigma, X' \sigma'} =  \bra{\Phi_{X\sigma}} \hat H_e  \ket{\Phi_{X'\sigma'}}$, and  there are 16 elements in total. Since the DQD basis states are linear combinations of the local dot states, these matrix elements are also linear combinations of the Hamiltonian matrix elements with respect to the local dot states,
$\wt{T}_{X\sigma, X' \sigma'} =  \bra{\phi_{X\sigma}} \hat H_e  \ket{\phi_{X'\sigma'}}$, by the basis transformation given by  \Eqref{eq:DQD-basis}.
According the \Eqref{eq:local-pot}, we can decompose $H_e=H_\rL + \Delta V_\rL=H_\rR + \Delta V_\rR$, and calculate 
\begin{equation}
 \renewcommand{\arraystretch}{1.5}
\wt{T} =
\begin{pmatrix}
  \varepsilon_{\rL\downarrow}{ +} v_{-} & 0 & s_d (\varepsilon_{\rR\downarrow} {+} v_0) & s_x(\varepsilon_{\rR\uparrow}{+}v_0) \\
  0 &   \varepsilon_{\rL\uparrow} {+} v_{-}  &s_x (\varepsilon_{\rR\downarrow}{+}v_0)  & s_d^*(  \varepsilon_{\rR\uparrow} {+}v_0) \\
  s_d^* (\varepsilon_{\rR\downarrow} {+} v_0^*) & s_x^* (\varepsilon_{\rR\downarrow}{+}v_0^*) &   \varepsilon_{\rR\downarrow}{ +} v_{+}  & 0 \\
s_x^* (\varepsilon_{\rR\uparrow}{+}v_0^*) &  s_d (\varepsilon_{\rR\uparrow}{ + }v_0^*) & 0 &  \varepsilon_{\rR\uparrow} {+} v_{+} 
 \end{pmatrix}
\end{equation}
where $\varepsilon_{\rL/\rR,\sigma}=\pm \varepsilon+\ss \tfrac{1}{2} \varepsilon_{\rL/\rR,\rZ}$ is the energy of the local Hamiltonian $H_{\rL/\rR}$,
while
\begin{equation}
 \begin{aligned}
 v_{-} & \equiv 
  \matele{0}{\ee^{- \ii pd } \Delta V_{\rL} \ee^{\ii pd }}{0}
  =  \matele{0}{\Delta V_{\rL}(x-d) }{0},\\
   v_{+} &\equiv
    \matele{0}{\ee^{ \ii pd } \Delta V_{\rR} \ee^{-\ii pd }}{0}=  \matele{0}{\Delta V_{\rR}(x+d) }{0},\\
 \end{aligned}
\end{equation}
are the energy corrections due to the existence of the other dot, and finally, 
\begin{equation}
 v_0  \equiv
 \frac{\matele{0}{\ee^{-\ii pd} \Delta V_\rR \ee^{-\ii pd}}{0}}{\matele{0}{\ee^{-2\ii  pd}}{0}} = \bra{0} \Delta V_\rR(x) \ket{0},
\end{equation}
where we have used the Wick's theorem to simplify the vacuum expectations (see Appendix \ref{app:v0}). 
Since $\Delta V_{\rL/\rR}$ is by definition close to zero in the vicinity of $x=\mp d$, the values of $v_{-}$ and $v_{+}$ are typically much smaller than other energy scales.  
On the other hand,  $v_0$ is related to barrier of the DQD system.
Specifically, for $d\gg x_0$, one can approximate 
\begin{equation}
\begin{aligned}
  v_0 \approx  V(0) -V(d)- \tfrac{1}{2}m^\star_e \Omega^2 d^2.
\end{aligned}
\end{equation}
As illustrated in \Figref{fig:DQD-pot-v0}, the parabolically shaped $V_\rR$ will tend to overestimate the DQD potential $V$ at $x=0$. Therefore $v_0$ is negative and its magnitude is typically much greater than other energy scales such as the orbital detuning and the Zeeman energy.  Furthermore, increasing the barrier height $V_{\mathrm{ba}}$ will decrease $\abs{v_0}$, thus decreasing the tunneling amplitude. This is consistent with typical observations. 

Transformating to the DQD basis and keep the terms up to the first order of the smallness factor $\ee^{-(d/x_0)^2}$, we find,
\begin{equation}
 \renewcommand{\arraystretch}{1.5}
T = \begin{pmatrix}
  \varepsilon_{\rL\downarrow}+ v_{-} & 0 &
 t_{\Downarrow}  &  s_{\Downarrow}\\
  0 &   \varepsilon_{\rL\uparrow} + v_{-}  &
s_{\Uparrow} &t_{\Uparrow} \\
t_{\Downarrow}^* & s_{\Uparrow}^* &   
 \varepsilon_{\rR\downarrow} + v_{+}  & 0 \\
s_{\Downarrow}^* & t_{\Uparrow}^* & 
0 &  \varepsilon_{\rR\uparrow} + v_{+} 
 \end{pmatrix},
\end{equation}
where we introduce the tunneling coefficients
\begin{equation}\label{eq:spin-tunnel-coeffs}
\begin{aligned}
t_\Downarrow & = s_d \left(
\tfrac{1}{2} \varepsilon_{\rR\downarrow}-\tfrac{1}{2} \varepsilon_{\rL\downarrow}
+v_0 -\tfrac{1}{2} v_{-} -\tfrac{1}{2} v_{+}
\right),\\
t_\Uparrow & = s_d^* \left(
\tfrac{1}{2} \varepsilon_{\rR\uparrow}-\tfrac{1}{2} \varepsilon_{\rL\uparrow}
+v_0 -\tfrac{1}{2} v_{-} -\tfrac{1}{2} v_{+}
\right),\\
s_\Downarrow & = s_x \left(\tfrac{1}{2} \varepsilon_{\rR\uparrow}-\tfrac{1}{2} \varepsilon_{\rL\downarrow}
+v_0 -\tfrac{1}{2} v_{-} -\tfrac{1}{2} v_{+}\right), \\
s_\Uparrow & = s_x \left(\tfrac{1}{2} \varepsilon_{\rR\downarrow}-\tfrac{1}{2} \varepsilon_{\rL\uparrow}
+v_0 -\tfrac{1}{2} v_{-} -\tfrac{1}{2} v_{+}\right). \\
\end{aligned}
\end{equation}

The dot Hamiltonian $H_d$ is defined according to the diagonal part of the $T$ tensor. We may redefine the effective chemical potentials to $\wt\varepsilon_{\rL\sigma}=\varepsilon_{\rL\sigma}+v_{-}$ and $\wt\varepsilon_{\rR\sigma}=\varepsilon_{\rL\sigma}+v_{+}$; or, as argued earlier, simply neglect the $v_{\pm}$ terms since they are very small.
Since shifting the global energy level has trivial influence on the qubit dynamics,
We can reset $\wt\varepsilon_L +\wt\varepsilon_R =0$ and define detuning $\varepsilon = \wt\varepsilon_\rL-\wt\varepsilon_\rR$.
 by
\begin{equation}
 \hat H_d = \sum_{\sigma=\Uparrow,\Downarrow} \tfrac{1}{2}  (\varepsilon +\ss \varepsilon_{\rZ,\rL}) n_{\rL \sigma} +\tfrac{1}{2}  (-\varepsilon +\ss \varepsilon_{\rZ,\rR}) n_{\rR \sigma}
\end{equation}
The Zeeman splitting energy can be further rearranged into the average splitting $\varepsilon_\rZ = (\varepsilon_{\rZ,\rL} + \varepsilon_{\rZ,\rR})/2$ and the energy difference $\delta\varepsilon_\rZ = (\varepsilon_{\rZ,\rL} - \varepsilon_{\rZ,\rR})$.

The tunneling Hamiltonian  $H_t$ is the off-diagonal part of the $T$ tensor.
In principle, there are mechanisms involved: spin-preserved and spin-flipped tunneling, in addition to on-site spin flipping. If there is no oscillatory electric field, on-site flipping term is absent and we have, 
\begin{equation}
    H_t = \sum_{\sigma=\Uparrow,\Downarrow} 
    (t_\sigma \mkern1mu a^{\sdagger}_{\rL\sigma} a^\phsdag _{\rR\sigma}+ s_{\sigma}  
    a^{\mathsmaller{\dagger}}_{\rL\sigma} a^\phsdag_{\rR\overline  \sigma} + h.c.),
\end{equation}
where spin-dependent tunneling coefficients are given by \Eqref{eq:spin-tunnel-coeffs}.
Calculating these coefficients requires knowledge of the full potential profile, but the details of the potential contributes trivially to the the gate dynamics by $v_0,v_{-}$ and $v_{+}$. 
In the presence of the barrier term $v_0$, the Zeeman energy 
differences in the parentheses of \Eqref{eq:spin-tunnel-coeffs} can be safely ignored. 
Combining with \Eqref{eq:wavefun-overlaps}, we can define a common tunneling strength $t_0\propto  \ee^{-(d/x_0)^2} $ for the spin-conserved tunneling $t$ and spin-flipped tunneling $s$, which are related by
\begin{align}
 t\equiv t_{\Downarrow} &= t_{\Uparrow}^* =t_0 \Bigl[\cos(\gamma_{\so}) - \ii \sin(\gamma_{\so}) \cos(\vartheta)\Bigr],\\
 s \equiv  s_{\Downarrow} &= s_{\Uparrow} = t_0\Bigl[ \phantom{\cos(\gamma_{\so})}-\ii \sin(\gamma_{\so})\sin(\vartheta)\Bigr].
\end{align}

Finally, the electron interaction energy involves in many terms and can also contribute to the exchange interaction. This is known as direct exchange and it mainly affects doubly occupied sites. This differs from the kinetic exchange that arises from $H_t$. Here we only consider leading order effect where all terms in the $V$ tensor are ignored and only on-site Coulomb repulsion is considered.  This can be summarized as 
\begin{equation}
 H_C = \sum_{j=\rL,\rR} \sum_{\sigma=\Uparrow,\Downarrow} \frac{U}{2}  a^\dagger_{j\sigma} a^\dagger_{j \sigma}  a_{j\bar \sigma} a_{j\bar \sigma}
\end{equation}

\subsubsection{Deriving the electro-static potential terms}\label{app:v0}

We  have derived the expression of $v_0$ in the following manner. Define 
$ \alpha_d ={d}/{\sqrt{2}x_0}$, then 
\begin{equation}
\ee^{-\ii p d }\ket{0} = D(\alpha_d)\ket{0} = \ket{\alpha_d},
\end{equation}
where $D(\alpha)$ is the displacement operator and $\ket{\alpha}$ represents a coherent state. The smallness factor considered in our problem is essentially
\begin{equation}
  \bra{0} D(\alpha_d) D(\alpha_d) \ket{0} = \braket{-\alpha_d}{\alpha_d}=\ee^{-(d/x_0)^2},
\end{equation}
and we can rewrite $v_0$ as 
\begin{equation}
 v_0 = \frac{\bra{-\alpha_d} \Delta V_\rR \ket{\alpha_d} }{ \braket{-\alpha_d}{\alpha_d}}.
\end{equation}
To calculate the numerator, we consider expanding it as a power series of $x$. This would require calculation of the matrix element:
\begin{equation}
 \bra{-\alpha_d} x^n \ket{\alpha_d} = \left(\frac{x_0}{\sqrt{2}}\right)^n \bra{-\alpha_d} (a+a^\dag)^n \ket{\alpha_d} 
\end{equation}
Expanding the $n$th power will lead to a summation of all possible sequences of $n$ field operators.
Using Wick's theorem, any field operator sequence can be converted to the normal order sequence plus all possible normal ordered contractions. Therefore,
\begin{equation}
 \begin{aligned}
 (a+a^\dag)^n  & = \sum_{k=0}^n\binom{n}{k}(a^\dagger)^k  a^{n-k} + \sum_{\text{1-contra.}}\sum_{k=0}^{n-2} \binom{n-2}{k}(a^\dagger)^k  a^{n-2-k} \\
 &+ \sum_{\text{2-contra.}}\sum_{k=0}^{n-4}\binom{n-4}{k}(a^\dagger)^k  a^{n-4-k} 
 + \cdots
 \end{aligned}
\end{equation}
 Using the property $a\ket{\alpha_d}=\alpha_d \ket{\alpha_d}$ and its conjugate variant, we have 
\begin{equation}
 \bra{-\alpha_d}  \sum_{k=0}^m\binom{m}{k}(a^\dagger)^k  a^{m-k} \ket{\alpha_d} =  \sum_{k=0}^m\binom{m}{k}(-\alpha_d)^k  {\alpha_d}^{m-k}=0.
\end{equation}
The only nonzero terms would be the fully contracted terms, which requires $n$ to be even and there are $(n-1)!!$  possible contractions. 
Apart from the factor of $\braket{-\alpha_d}{\alpha_d}$, this result coincides with the vacuum expectation value. Hence
\begin{equation}
 v_0 = \bra{0} \Delta V_\rR(x) \ket{0}.
\end{equation}

\subsection{Hamiltonian in the computational space}
  In order to derive the effective exchange Hamiltonian, we expand the system Hamiltonian in the six low energy  two-electron  spin basis that consists of 4 states in the $(1,1)$ configuration and the singlet state $S(2,0)$ and $S(0,2)$.
 Each of these basis are antisymmetric two-electron wave functions.
The  tunneling Hamiltonian, for example, acts on the $\ket{\downarrow\uparrow}$ by 
\begin{equation}
\begin{split}
 H_t \ket{\downarrow\uparrow} &= H_t \frac{1}{\sqrt{2}} \left(\, 
   \ket{\Phi_{\rL \Downarrow}}  \ket{\Phi_{\rR \Uparrow}} - 
     \ket{\Phi_{\rR \Uparrow}}\ket{\Phi_{\rL \Downarrow}}\right)\\
 &= - t^* \ket{S(2,0)} -  t^*  \ket{S(0,2)}.
\end{split}
 \end{equation}
Using similar calculations, we can represent the low energy Hamiltonian as
\begin{equation}\label{eq:lowE-H6}
 \renewcommand{\arraystretch}{1.5}
H_{6}\,\widehat{=}\,
\left(
\begin{array}{cccccc}
\varepsilon_\rZ& 0 & 0 & 0 & s^* & -s \\
 0 & \tfrac{1}{2}\delta\varepsilon_\rZ & 0 & 0 & t^* & t^* \\
 0 & 0 & -\tfrac{1}{2}\delta\varepsilon_\rZ & 0 & -t & -t \\
 0 & 0 & 0 & -\varepsilon_\rZ & -s^* & s \\
 s & t & -t^* & -s & U-\varepsilon  & 0 \\
 -s^* & t & -t^* & s^* & 0 & U+\varepsilon  \\
\end{array}
\right),
 \end{equation}
 under the basis $\{\ket{\uparrow \uparrow},
\ket{\uparrow \downarrow},
\ket{\downarrow \uparrow},
\ket{\downarrow \downarrow},
\ket{S(2,0)},
\ket{S(0,2)}\}$.
Assuming that the on-site Coulomb energy is much larger than the tunneling energy:
$U\gg t_0$, we would have three different energy blocks---$(1,1),(0,2),(2,0)$ with weak coupling. The DC gates work exclusively in the $(1,1)$ charge configuration. To find the influence of other energy levels on the $(1,1)$ space, we project $H_{6}$ in to four $(1,1)$ states space using the Schrieffer-Wolff transformation
$ e^{S} H_{6} e^{-S} $.
 Define the diagonal and off-diagonal part of $H_6$ as $H_{60}$ and $H_{61}$, 
the transformation matrix $S$ must satisfy the condition
$ [H_{60}, S] = H_{61}$. And the effective Hamiltonian in the $(1,1)$ charge configuration can be approximate with 
 \begin{align}\label{eq:SWtransform}
H= \mathcal P_{(1,1)} \left( H_{60} + \frac{1}{2}[S, H_{61}] \right) +O((t_0/U)^4).
 \end{align}
Since $H_{60}$ is diagonal, $S$ can be explicitly found by,
 \begin{align}
S_{ii}=0,\quad S_{ij} = \frac{(H_{61})_{ij}}{(H_{60})_{ii}- (H_{60})_{jj}} \quad (i\neq j).
 \end{align}
Substituting into \Eqref{eq:SWtransform}, we can write the resulting effective Hamiltonian as
\begin{equation}
 H = H_0 + H_\mathrm{ex},
\end{equation}
where, 
\begin{equation}
H_0 \,\widehat{=}\,  \diag( \varepsilon_\rZ,\tfrac{1}{2}\delta\varepsilon_\rZ,-\tfrac{1}{2}\delta\varepsilon_\rZ,- \varepsilon_\rZ)
\end{equation}
is a diagonal matrix that arises solely from the local Zeeman field for each dot.  $H_\mathrm{ex}$ is responsible for the exchange coupling, explicitly given by its $(i,j)$th element,
\begin{equation}
 (H_\mathrm{ex})_{ij} = -2(\mb j_i + \mb j_j) \bm \xi_i \bm \xi_j^*,
\end{equation}
where we introduce vectors 
$\mb j =  (j^a_-, j^b_-, j^b_+, j^a_+)^T$, and
$\bm{\xi}  = 1/(\sqrt{2}t_0)\, (s^*, t^*, -t, s)^T$, with 
\begin{equation}
 \begin{aligned}
 j^a_{\pm} & = \frac{t_0^2}{2} \left(\frac{1}{U\pm\varepsilon_\rZ-\varepsilon} + \frac{1}{U\pm\varepsilon_\rZ+\varepsilon}  \right), \\
  j^b_{\pm} & = \frac{t_0^2}{2} \left(\frac{1}{U\pm \delta\varepsilon_\rZ-\varepsilon} + \frac{1}{U\pm \delta\varepsilon_\rZ+\varepsilon}  \right).
\end{aligned}
\end{equation}
By assumption $U\gg \varepsilon_\rZ, \delta\varepsilon_\rZ$, we can make the approximation
\begin{align}\label{eq:H1}
 j^a_{\pm}\approx  j^b_{\pm}\approx \frac{t_0^2}{2} \left(\frac{1}{U-\varepsilon} + \frac{1}{U+\varepsilon}\right)\equiv \frac{J}{4}.
\end{align} 
Due to the form $\bm{\xi}$ is defined, we can define normalized state $\ket{\xi}$ as the entangled two-qubit state
\begin{equation}
 \ket{\xi} =\frac{1}{\sqrt{2}}  \left( \ket{\uparrow} \ket{\nearrow}- \ket{\downarrow} \ket{\swarrow}\right) ,
\end{equation}
where, 
\begin{equation}
\begin{cases}
  \ket{\nearrow} = \wt s^* \ket{\uparrow} + \wt t^* \ket{\downarrow},\\
    \ket{\swarrow} = \wt t \ket{\uparrow} -\wt s \ket{\downarrow}
\end{cases}
\end{equation}
are a pair of orthogonal spin states with normalized coefficients $\abs{
\wt s}^2+\abs{\wt t}^2=1$ given by $\wt s= s/t_0$ and $\wt t = t/t_0$.
This leads to the final expression for the exchange coupling Hamiltonian
\begin{equation}\label{eq:Hex}
 H_\mathrm{ex} = - J \ket{\xi}\bra{\xi},
\end{equation}
a very compact outcome.

\section{Perturbative results for the CZ gate}\label{app:CZpert}
Let us calculate the eigenenergies of $H$ to the second order. Using non-degenerate perturbation theory, we obtain 
\begin{equation}
 \begin{aligned}
\delta E_1 &\approx -\left(\frac{J}{\varepsilon_\rZ}\right) \frac{\abs{\wt s}^2}{2} + 
\left(\frac{J}{\varepsilon_\rZ}\right)^2 
\left( \frac{\abs{\wt s}^4}{8}+\frac{\abs{\wt s}^2 \abs{\wt t}^2}{4+2\delta} 
+\frac{J^2 \abs{\wt s}^2 \abs{\wt t}^2}{4-2\delta} \right),\\
\delta E_2 &\approx -\left(\frac{J}{\varepsilon_\rZ}\right) \frac{\abs{\wt t}^2}{2} + 
\left(\frac{J}{\varepsilon_\rZ}\right)^2 
\left( \frac{\abs{\wt t}^4}{4\delta}+\frac{\abs{\wt s}^2 \abs{\wt t}^2}{4+2\delta} 
-\frac{J^2 \abs{\wt s}^2 \abs{\wt t}^2}{4-2\delta} \right),\\
\delta E_3 &\approx -\left(\frac{J}{\varepsilon_\rZ}\right) \frac{\abs{\wt t}^2}{2} + 
\left(\frac{J}{\varepsilon_\rZ}\right)^2 
\left( -\frac{\abs{\wt t}^4}{4\delta}-\frac{\abs{\wt s}^2 \abs{\wt t}^2}{4+2\delta} 
+\frac{J^2 \abs{\wt s}^2 \abs{\wt t}^2}{4-2\delta} \right),  \\
\delta E_4 &\approx -\left(\frac{J}{\varepsilon_\rZ}\right) \frac{\abs{\wt s}^2}{2} + 
\left(\frac{J}{\varepsilon_\rZ}\right)^2 
\left( -\frac{\abs{\wt s}^4}{8}-\frac{\abs{\wt s}^2 \abs{\wt t}^2}{4+2\delta} 
-\frac{J^2 \abs{\wt s}^2 \abs{\wt t}^2}{4-2\delta} \right),
 \end{aligned}
\end{equation}
where $\delta\equiv\delta\varepsilon_\rZ/\varepsilon_\rZ$. The first and second order energy corrections all lead to the evolution time in \Eqref{eq:CZ-time}.
 Combined with the phase-matching conditions $\phi_0= -\delta E_1 \tau_\CZ$, 
 $\phi_0+ \phi_\rR=- \delta E_2 \tau_\CZ$ and  $\phi_0+\phi_\rL=- \delta E_3  \tau_\CZ$, the corresponding local phase corrections are found to be 
\begin{equation}
 \begin{aligned}
   \phi_\rL &=\pm\frac{\pi}{2} \left[1+\frac{J}{4\varepsilon_\rZ} 
   \frac{2 (2 + \delta) \abs{\wt t}^4+ \delta  (2 + \delta) \abs{\wt s}^4+8 \delta  \abs{\wt s}^2 \abs{\wt t}^2 }
   {4 \delta  (2 + \delta) (\abs{\wt t}^2- \abs{\wt s}^2)}\right],\\
   \phi_\rR  &= \pm \frac{\pi}{2} \left[1 - \frac{J}{4\varepsilon_\rZ} 
   \frac{2 (2- \delta) \abs{\wt t}^4 - \delta  (2- \delta) \abs{\wt s}^4-8 \delta  \abs{\wt s}^2 \abs{\wt t}^2 }
   {4 \delta  (2- \delta) (\abs{\wt t}^2- \abs{\wt s}^2)}\right],
 \end{aligned}
\end{equation}
where the $\pm$ signs in the front is `$+$' when $\abs{\wt{t}}>\abs{\wt s}$ and `$-$' when $\abs{\wt{t}}<\abs{\wt s}$. 

The global and local phase corrections define a phase correction vector 
$\vec\phi_\mathrm{corr}=\tau_\mathrm{CZ} (\delta E_1, \delta E_2, \delta E_3, \delta E_2+ \delta E_3 - \delta E_1)$.
Carrying out the phase corrections to $\wt U(\tau_\CZ)$ produces the phase-corrected gate $\wt U_\mathrm{c}(\tau_\CZ) =\ee^{\ii \phi_\mathrm{corr} }\wt U(\tau_\CZ) $, whose diagonal elements are given by  
\begin{equation}
\bra{n} \wt U_\mathrm{c}(\tau_\CZ) \ket{n} 
= \sum_m 
\abs{\braket{n}{m'}}^2 \exp[{-\ii (E_m' -E_n^{(0)}) \tau_\CZ +\ii \vec\phi_\mathrm{corr,n}}] \equiv  \sum_m 
\abs{\braket{n}{m'}}^2  \Phi_{n,m},
\end{equation}
where the phase matrix $\Phi_{n,m}$ introduced above is given by 
\begin{equation}
 \Phi_{n,m} = \begin{pmatrix}
  1 & \ee^{-\ii (1-\delta/2)\tau_\CZ} & \ee^{-\ii (1+\delta/2)\tau_\CZ} & \ee^{-\ii 2\tau_\CZ} \\
\ee^{\ii (1-\delta/2)\tau_\CZ} &1  & \ee^{-\ii \delta\tau_\CZ} & \ee^{-\ii (1+\delta/2)\tau_\CZ} \\
\ee^{\ii (1+\delta/2)\tau_\CZ} & \ee^{\ii \delta\tau_\CZ}  &1  & \ee^{-\ii (1-\delta/2)\tau_\CZ} \\
- \ee^{\ii 2\tau_\CZ} & - \ee^{\ii (1+\delta/2)\tau_\CZ}  & -  \ee^{\ii (1-\delta/2)\tau_\CZ}  & -1
 \end{pmatrix}.
\end{equation}
The perturbed eigenstates can be calculated as, 
\begin{equation}
 \begin{aligned}
\ket{1'} &=\frac{1}{\sqrt{N_1}}\left[ \ket{1} + \frac{J}{\varepsilon_\rZ} \Bigl( 
- \frac{s t^*}{2-\delta} \, \ket{2} + \frac{s t}{2+\delta}\,  \ket{3} - \frac{s^2}{4}\, \ket{4}\Bigr) + O\Bigl( \frac{J}{\varepsilon_\rZ} \Bigr)^2 \right] \\
\ket{2'} &=\frac{1}{\sqrt{N_2}}\left[ \ket{2} + \frac{J}{\varepsilon_\rZ} \Bigl( 
\frac{s^* t}{2-\delta} \, \ket{1} - \frac{s t}{2+\delta}\,  \ket{4} + \frac{t^2}{2\delta}\, \ket{3}  \Bigr) + O\Bigl( \frac{J}{\varepsilon_\rZ} \Bigr)^2 \right]\\
\ket{3'} &=\frac{1}{\sqrt{N_3}}\left[ \ket{3} + \frac{J}{\varepsilon_\rZ} \Bigl(\frac{s t^*}{2-\delta} \, \ket{4} - \frac{s^* t^*}{2+\delta}\,  \ket{1} - \frac{(t^*)^2}{2\delta}\, \ket{2}  \Bigr) + O\Bigl( \frac{J}{\varepsilon_\rZ} \Bigr)^2\right] \\
\ket{4'} &=\frac{1}{\sqrt{N_4}}\left[ \ket{4} + \frac{J}{\varepsilon_\rZ} \Bigl( - \frac{s^* t}{2-\delta} \, \ket{3} + \frac{s^* t^*}{2+\delta}\,  \ket{2} + \frac{(s^*)^2}{4}\, \ket{1}  \Bigr) + O\Bigl( \frac{J}{\varepsilon_\rZ} \Bigr)^2\right],
 \end{aligned}
\end{equation}
where the normalization factors $N_1$, $N_2$, $N_3$ and $N_4$ can be determined up to $O(J/\varepsilon_\rZ)^2$ from the above.  The gate infidelity  is calculated by taking the trace product of $\wt U_\mathrm{c}(\tau_\CZ)$ with the $\CZ$ gate, which only involves in the diagonal elements specified above. 
After substituting the evolution time and phase corrections, we can calculate, up to  $O(J/\varepsilon_\rZ)^2$, the gate infidelity as 
\begin{multline}
  \InF_{\CZ} = \frac{1}{20}  \Bigl(\frac{J}{\varepsilon_\rZ}\Bigr)^2  \Bigg(
\abs{\wt s}^4 \left[ 1- \cos(2\varepsilon_\rZ \tau_\CZ)\right] +
\frac{4\abs{\wt t}^4}{\delta^2}\! \left[1- \cos(\delta\varepsilon_\rZ \tau_\CZ)\right] \\ +
\frac{32 \abs{\wt s}^2 \abs{\wt t}^2}{(2-\delta)^2}  \left[ 1-\cos(\varepsilon_\rZ(1-\tfrac{\delta}{2}) \tau_\CZ) \right] 
+\frac{32 \abs{\wt s}^2 \abs{\wt t}^2}{(2+\delta)^2}  \left[1-\cos(\varepsilon_\rZ(1+\tfrac{\delta}{2}) \tau_\CZ)\right]\Bigg) + O\Bigl(\frac{J}{\varepsilon_\rZ}\Bigr)^3.
\end{multline}
This infidelity expression depends on the evolution time $\tau_\mathrm{CZ}$ through the cosine terms. We can further discard the oscillatory cosine parts and bound 
\begin{equation}
  \InF_{\CZ} \lesssim \frac{1}{10}  \Bigl(\frac{J}{\varepsilon_\rZ}\Bigr)^{\!2} \! \Bigg[
\abs{\wt s}^4  +\frac{4\abs{\wt t}^4}{\delta^2} 
+\frac{64(4+\delta^2) }{(4-\delta^2)^2}   \abs{\wt s}^2 \abs{\wt t}^2\Bigg].
\end{equation}

\end{document}